\documentclass[11pt,prb]{revtex4} 
\usepackage{amsfonts}
\usepackage{amsmath}
\usepackage{amssymb}
\usepackage{graphicx}
\usepackage{hyperref}

\providecommand\bcdot{\boldsymbol{\cdot}}
\providecommand\btau{\mbox{\boldmath $\tau$}}
\providecommand\bgam{\mbox{\boldmath $\gamma$}}

\begin{document}

\title{Viscoelastic flow transitions in abrupt planar contractions}
\author{Lars H. Genieser$^{1}$, Arvind Gopinath$^{2}$, Robert C. Armstrong$^{1}$, and
Robert A. Brown$^{1}$}
\affiliation{ $^{1}$Department of Chemical Engineering, Massachusetts
Institute of Technology, Cambridge, MA, USA. \\
$^{2}$Fisher School of Physics, Brandeis University, Waltham, MA, USA.}
\date{January, 2004}%

\maketitle

\section*{ABSTRACT}

We present experimental evidence of global viscoelastic flow
transitions in 2:1, 8:1 and 32:1 planar contractions under inertia-less conditions.
Light
sheet visualization and laser Doppler velocimetry techniques
are used to probe
spatial structure and time scales associated with the onset of these
instabilities. The results are reported in terms of
critical Weissenberg numbers characterizing the fluid flow rates. For a given contraction ratio
and polymer fluid, a two-dimensional, steady flow with
converging streamlines transitions to a two-dimensional pattern with
diverging streamlines beyond a critical
Weissenberg number. At even higher Weissenberg numbers, spatial
transition to three-dimensional flow is observed.
A relationship between the upstream
Weissenberg number for the onset of this spatial instability and the
contraction ratio is derived.
For contraction ratios
substantially greater than unity, we observe that further increase in the flow rate
results in a second temporal transition
to a three-dimensional, time-dependent flow. This complex flow
arises due to a combination of the effects of stress-curvature
interaction and three dimensional perturbations induced by the walls
bounding the neutral direction. Comparison
with studies on other geometries indicates that boundary shapes
deeply influence the sequence and nature of flow transitions.

{\em Keywords} Viscoelastic flow transitions, planar contraction
flow, critical Weissenberg number

\section{Introduction}

A common and important class of flows is constituted by the motion
of a polymer liquid through abrupt planar and axisymmetric
contraction geometries. These flows also constitute canonical
benchmarks to test numerical codes and hence have been a
subject of a few experimental studies. In this article,
we describe detailed experimental studies on viscoelastic
flow transitions in 2:1, 3:1 and 32:1 planar contraction geometries.
A novel Boger fluid that allows us to obtain flows with negligible inertial
effects is used as the test fluid, thus the instabilities are purely
due to fluid elasticity. Our focus is on the
classification of global flow transitions in the planar contraction,
i.e., those which occur over the longest characteristic length scale
of the geometry which is the upstream half-height.
The flow transitions are characterized via a suitably defined aspect ratio that captures
the effects of flow geometry, and the Weissenberg number ($We = \lambda
/t_{\mathrm{def}}$),
$\lambda$ being a characteristic relaxation time of the
viscoelastic fluid and $t_{\mathrm{def}}$ a
characteristic measure of the local deformation time scale (an
inverse strain rate) respectively.

Previous investigations of complex flow through diverse geometries
[1-13] have revealed transitions in the flow structure
associated with viscoelastic effects as the flow rate is varied.
Binding and Walters [2] used
a polyacrylamide-based Boger fluid in studies on flows through a
14:1 planar contraction at flow conditions for which inertial
effects were in-significant. At high flow rates, a two-dimensional
rearrangement of the flow field to a pattern of diverging
streamlines occurred upstream of the contraction plane. This pattern
was also observed in flow experiments by Evans and Walters [3] of a
polyacrylamide solution through a 4:1 contraction under conditions
where inertial effects could not be neglected. Chiba et al. [4,6]
used non-dilute shear-thinning, aqueous polyacrylamide solutions
flowing through planar contractions. For contraction ratios of 3.3:1
and 5:1 but not for the 10:1 contraction, the onset of diverging flow was seen at a critical flow
rate. As in the
Evans and Walters study, inertia was a significant factor.
Similar rearrangement of the flow is observed for viscoelastic
flows through axisymmetric contraction geometries. Experiments indicate
that lower contraction ratio geometries in both the
planar and axisymmetric contractions result in more pronounced diverging
flow behavior than higher contraction ratio configurations. The maximum contraction
ratio for which diverging flow is observed is also typically greater
for planar than for axisymmetric contraction flows

Transitions from a two-dimensional base flow to three-dimensional
steady or time-dependent flow have also been observed in flow
through abrupt contractions and also in flows through other
geometries.  Work by  Binding and Walters [2]
revealed transition to an asymmetry in the velocity field at a
critical $We$ greater than that associated with the transition to
diverging flow streamlines. The spatial structure of the post-transition
flow  comprised of vortical structures. Experiments by Chiba et al.
[4,6] in 3.3:1 and 10:1 contractions also revealed a transition from
two- to three-dimensional flows at flow rates above that required
for the onset of diverging flow. The spatial structure of the flow in these
experiments also
was found to consist of interlaced pairs of counter-rotating
vortices on each side of the center-plane. At yet higher flow rates, a
transition was seen from the three-dimensional, steady flow to
three-dimensional, time-dependent flow.
It was difficult to identify unambiguously the mechanism
driving the flow transition, as inertial effects were significant at
the flow rates attained.

In general velocity-field transition sequences
for a variety of complex flows
are evidenced to share a number of common features.
For a given
contraction ratio we expect the rheology of the
polymer fluid and elongational flow material functions to 
influence the evolution of the velocity field.
In all cases for
which a two-dimensional flow rearrangement occurred, the critical
$We$ was lower than the value of $We$ characterizing transition to
three-dimensional and/or time-dependent flow. The spatial structure
of the flow after transition had the form of G\"{o}rtler-type
vortices. However, there are also significant differences.
In some geometries, a direct transition from a
two-dimensional, steady flow to a three-dimensional, time-dependent
flow was observed; flows through other geometries first undergo a
spatial transition to three-dimensional flow and then a temporal
transition to time-dependent flow.

In the following sections, we report on 
extensive experimental results on the structure of the
viscoelastic flow transitions in planar contraction flow where both
contraction ratio and the flow rate characterized by a typical
Weissenberg number are varied.
Light sheet visualization
coupled with laser Doppler velocimetry (LDV) measurements is used to
characterize quantitatively the critical Weissenberg numbers,
length, and time scales associated with the onset of the
instabilities.
Scaling based on the viscoelastic
G\"{o}rtler number analysis complemented by plots of appropriate
measures of the velocity patterns are used to construct transition
and bifurcation maps.  Finally, the
flow structures observed in planar contraction flows are compared
with those identified in other flow geometries.

\section{Experimental test geometry and fluid handling system}

\begin{figure}
\includegraphics[width=14 cm]{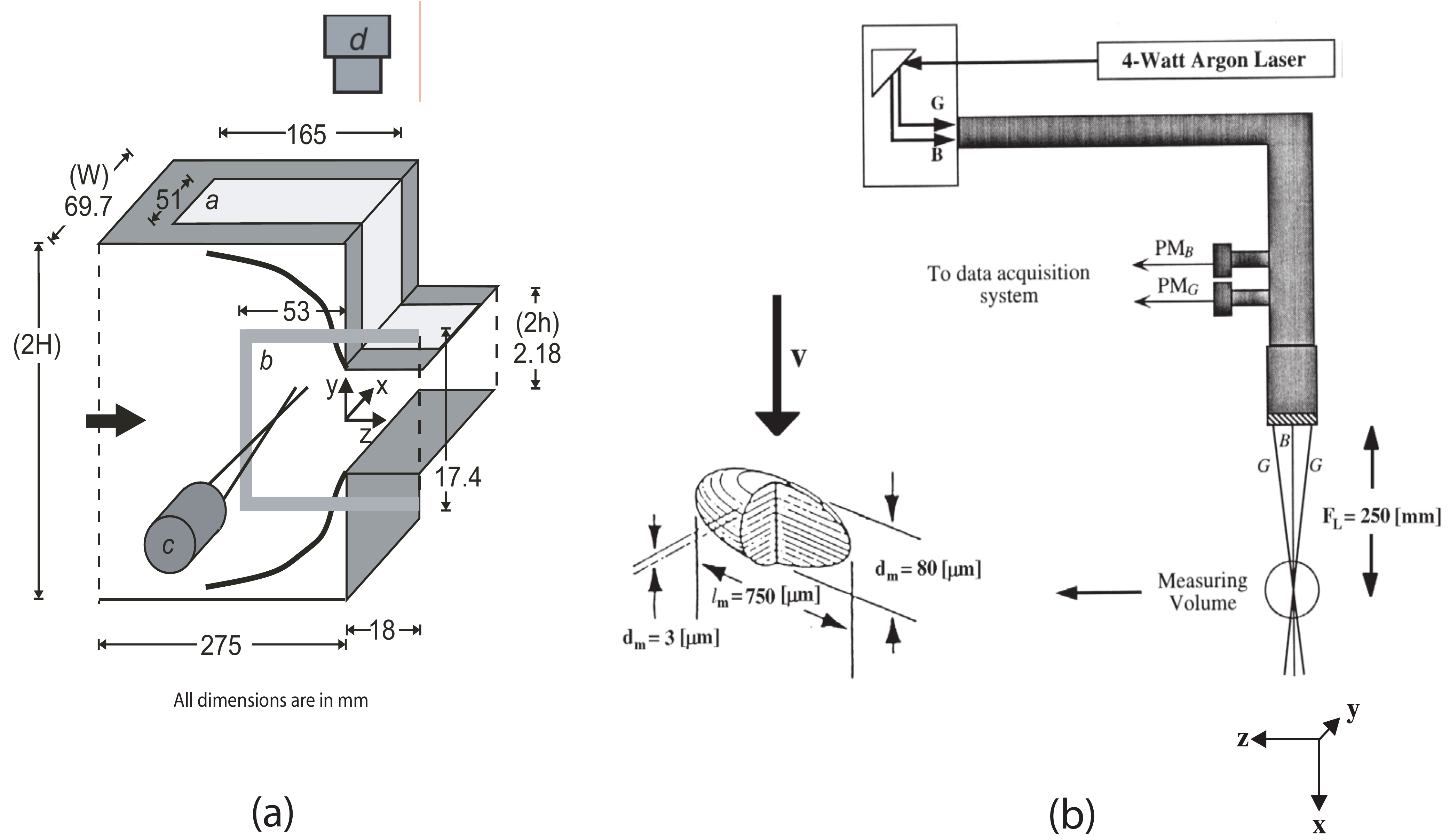}\
\caption { (A) Planar contraction test geometry. Fixed dimensions of the
geometry, $W$ and $2h$, and the dimensions of the windows are shown;
the upstream channel height, $2H$ is variable. (a) Top window for
flow visualization in the x-z plane, (b) The side window,
constructed of SF-57 glass, is used for LDV and FIB measurements.
(c) The final focusing optics from which the dual beams of the LDV
system emanate are illustrated; the FIB probe beam is not shown and
(d) video camera for recording images in the x-z plane (light
sheet). All dimensions are in millimeters. (B)
Schematic diagram of the two-component LDV system used in
this study and the measuring volume.}
\end{figure}

The planar contraction test geometry used in this study is
illustrated in Figure 1(A).
The exterior of the geometry was constructed of anodized aluminum.
Fluid flowed in
the $z$-direction from an upstream duct of half-height $H$ into a
smaller, downstream duct of half-height $h$. The width, $W$, is
fixed throughout the geometry. A downstream insert constructed of
anodized aluminum and borosilicate glass (BK-10, Schott Glass
Technologies) was used to set the half-height of the downstream
slit, $h = 1.09 $$\pm$ $0.05$ mm.
The contraction ratio, $H/h$, was varied by
changing the upstream insert also constructed of anodized aluminum and
polymethylmethacrylate (PMMA). Qualitative, full-field streakline
observations of the kinematic structure of the flow in the
$xz$-plane were made through windows placed in the top of the
geometry, $a$. Quantitative LDV measurements of the $v_{y}$ or
$v_{z}$ velocity component and full field streakline observations of
flow in the $yz$-plane were made through a window placed in the side
of the contraction geometry, $b$. The FIB measurements also were
conducted through the window, $b$. To minimize the influence of
parasitic birefringence on the FIB measurement, the window was
constructed of SF-57 glass (Schott Glass Technologies); this
material has a low stress optical coefficient of C = 0.02 $\times$
10$^{-12}$ Pa$^{-1}$ at 589 nm. The faces of the glass were polished
and coated to minimize reflections.
The fluid was driven in batch mode by nitrogen pressure from a supply tank
through the test cell  and into a receiving tank  maintained at atmospheric
pressure with the fluid flow rate adjusted by controlling the gas
pressure with a self-relieving regulator.

The coordinate system used throughout this paper is as indicated in Figure 1(A).
The origin is located in the center of the downstream duct, at the
contraction plane ($z = 0$, where the upstream and downstream ducts
join). The dimensional coordinates $(x, y, z)$ are given in
millimeters; dimensionless coordinates are based on the downstream
half-height: $\chi = x/h$, $\upsilon = y/h$, and $\zeta = z/h$. The
term center-plane refers to the plane defined by $(\chi, \upsilon=
0, \zeta)$; centerline refers to any line in this plane with a
constant value of $\chi$. Because the geometry is nominally
two-dimensional, $x$ is referred to as the neutral direction. The
width $W$ of the geometry was held constant at 69.8 mm, so that in
the upstream region, a truly two-dimensional flow was most closely
approximated for low contraction ratios. For the $H/h$ = 32
contraction ratio, the upstream aspect ratio was only $W/2H$ = 1.

The light sheet visualization technique was used to record a
streakline image of the velocity field in a selected two-dimensional
plane. A laser beam is passed through a cylindrical lens to form a
light sheet with thickness of approximately 100 $\mu m$ throughout
the illuminated region of the flow field. As the particles in the
fluid travel through the sheet they scatter light which is recorded
by a video camera; the axis of the video camera is normal to the
light sheet. Two configurations were used in the experiments: in the
first the light sheet was formed in the $xz$-plane to acquire a top
view.  In the second
the light sheet was formed in the $yz$-plane to acquire a side view.
The video camera signal was stored and subsequently
digitized using a frame grabber board. Frames in
a time series were superimposed and averaged using image processing
software (NIH Image v. 1.55). By averaging together several frames
separated by equal intervals of time streak-line images were
produced. The length and direction of a given streak corresponded to
the local velocity vector in the plane of the light sheet.
To visualize the three-dimensional spatial structure
of the instability, images were acquired for a given flow with light
sheets in the $xz$-plane at several $y$-positions and in the
$yz$-plane at various $x$-positions. These sets of two-dimensional
images were used to construct a three-dimensional schematic picture
of the flow field.

The specific configuration of the LDV system (TSI, Model 9100-12)
used is  illustrated in Figure 1(B). The output of a 4 Watt multi-line
argon-ion laser is passed through a series of optical elements and a
final focusing lens of focal length F$_{L}$ = 250 mm to form two
pairs of intersecting beams orthogonal to each other and capable of
measuring the $v_{y}$ (blue beam pair) and $v_{z}$ (green beam pair)
velocity components. The half-angle included by each beam pair is
0.083 rad; the half-angle in conjunction with the wavelength sets
the fringe spacing $d_{f}$ = 3.1 $\mu$m. The superimposed
measuring volumes from the two beam pairs are ellipsoids with dimension
80 $\mu$m $\times$ 80 $\mu$m $\times$ 500 $\mu$m in air, the long axis
positioned along the x-direction (in the 0.30 wt\% PIB in PB test
fluid, which has a relative refractive index of 1.50, the dimensions
of the measuring volume are approximately 80 x 80 x 750 $\mu$m). The
LDV optics are mounted on a three axis translating table (TSI, Model
9500) capable of positioning the measuring volume to within $\pm$ 4
$\mu$m.

Under steady flow conditions, velocity data is collected by
operating the system in the spectrum analysis mode. The Doppler
burst signals detected by the photomultiplier (PM) tube are passed
through an FFT spectrum analyzer (Nicolet, Model 660B) which
calculates the power spectrum (PS). The velocity of the particles
passing through the measuring volume is then determined from the
characteristic frequency of the peak in the PS. The PS of a number
of successive bursts is averaged together to enable accurate
measurement of low velocities.
For time-dependent flows, we use a frequency tracker
(DISA, Model 55N20/21) to follow the Doppler frequency.
To ensure adequate data collection rate, the fluid
was seeded with 2 $\mu$m silicon carbide scattering particles (TSI
10081); a seeding density of 0.036 g/l was used. This
ameliorated the problem of tracker drop out. 

\section{Determination of fluid properties}

The two-component experimental test fluid consisted of a homogeneous mixture of high molecular
weight polyisobutylene (PIB) (Exxon Vistanex L120), with
molecular weight of $M_w$ $= 2$ $\times$ 10$^{6}$ gmol$^{-1}$,
and polybutene (PB) (Amoco
Panalane H300E), with $M_w$ $\approx  1.3$ $\times$ 10$^{3}$
gmol$^{-1}$. The fluid was produced by first dissolving
polyisobutylene in chromatography grade hexane to a concentration of
2 wt \%; this solution was then combined with PB. The hexane was
evaporated under nitrogen purge and then under vacuum at 70C from
the PIB/PB/hexane mixture to obtain the 0.30 wt\% PIB in PB
solution. The two-component Boger test fluid did not exhibit beam
divergence when sheared in a Couette cell.

In order to interpret the experimental data, we considered both the
four-mode linear Maxwell model and the four-mode non-linear Giesekus
model. The generalized linear Maxwell model was used to interpret
the linear viscoelastic spectrum of the fluid whereas the Giesekus
model was used to characterize the shear thinning occurring at
finite strains. The multi mode formulation yields the total
stress tensor, $\btau$,
\begin{equation}
{\btau}(t) = \sum_{k=1}^{N_{\mathrm{modes}}} \btau_{k}(t) +
\btau_{s} =  \sum_{k=1}^{N_{\mathrm{modes}}} \btau_{k}(t)
- \eta_{s} \dot{\bgam}
\end{equation}
as a linear superposition of the
components associated with the $k$-th mode of the relaxation spectrum.
Since the PB solvent exhibits a Newtonian response on the time scale
and for the stress levels attained in these experiments, we
include the contribution of a Newtonian solvent
to the total stress in Equation (1). Linear
viscoelasticity is characterized using the generalized linear
Maxwell model for which the $k$-th mode can be written as
\[
\btau_{k} + \lambda_{k}{{\partial{\btau_{k}} \over \partial{t}}} = -
\eta_{k} \dot{\bgam}
\]
where $\dot{\bgam}$ is the
strain-rate tensor, $\lambda_{k}$, the relaxation time, $\eta_{k}$,
the viscosity, and $t$, the time. To describe the shear-thinning
nature of the fluid, we use the non-linear Giesekus model with the
$k$-th mode stress contribution given by
\[
\btau_{k} + \lambda_{k}({\btau_{k}})_1 -\alpha_{k}{\lambda_{k} \over
\eta_{k}} (\btau_{k} \bcdot \btau_{k}) = - \eta_{k} \dot{\bgam}.
\]

The temperature dependence of the material functions for the 0.30
wt\% PIB in PB test fluid was determined by measuring the dynamic
viscosity of a fluid sample undergoing oscillatory shear in a
cone-and-plate configuration and sweeping the temperature over a
range from 15C to 35C. An Arrhenius expression was used to
estimate the temperature dependence for the test fluid and pure
solvent. The rheology of the polybutene solvent was characterized in
a RMS-800 mechanical spectrometer. No decreasing trend in the
dynamic viscosity with increasing frequency was discerned up to
$\omega_{\mathrm{max}} = 100$ s$^{-1}$, the value up to this
frequency  being $\eta'_{s} = 79$ Pa s. This compared very well well
with the viscosity of the polybutene measured in steady-shear flow,
$\eta_{s}$ up to a shear rate of 26 s$^{-1}$. At greater shear rates
a gradual decrease in the viscosity with shear rate is apparent;
this was attributed to viscous heating.

\begin{figure}
\includegraphics[width=15cm]{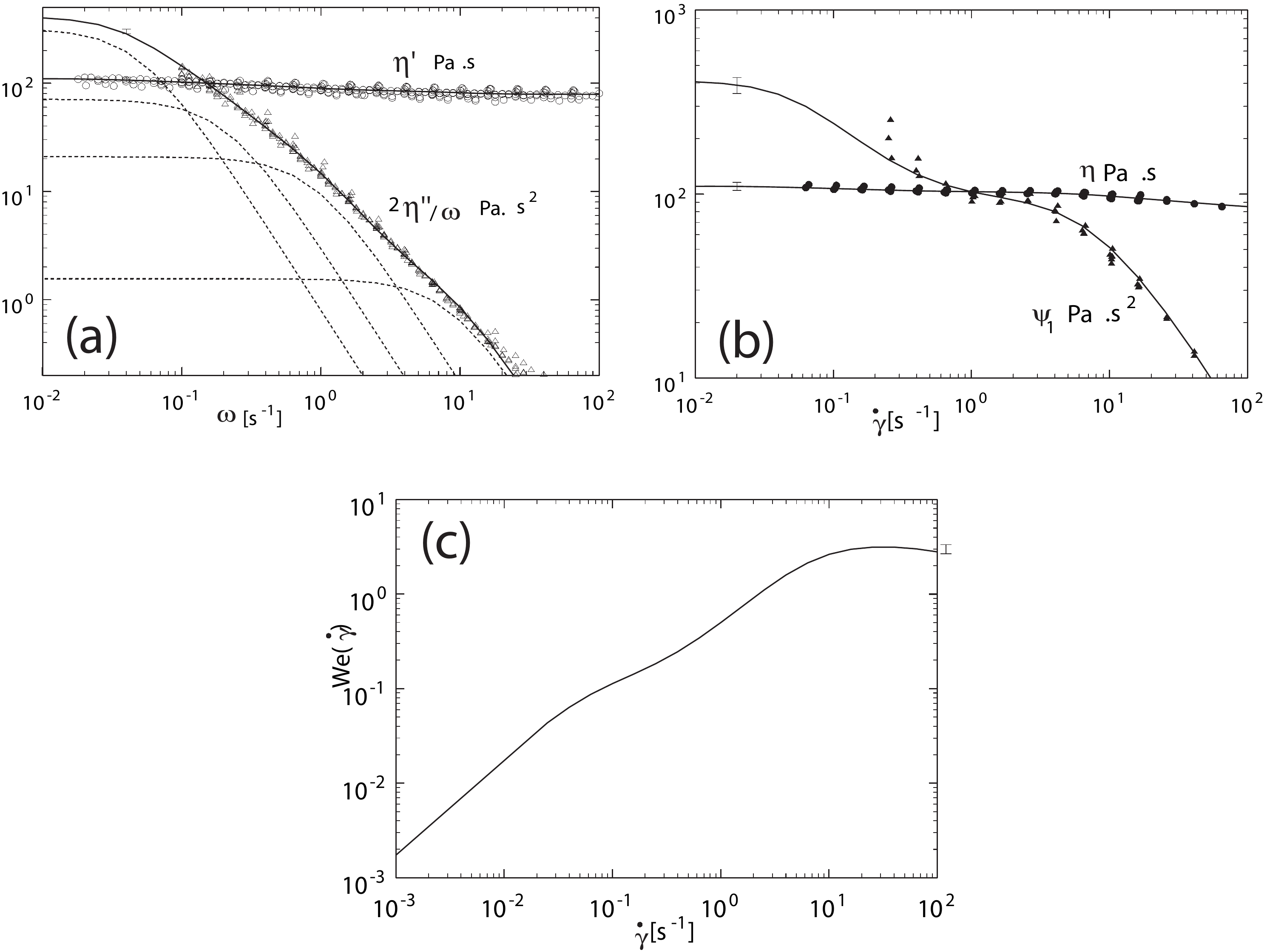}\
\caption{(a) Linear viscoelastic properties $\eta'(\omega)$ and
$2\eta''(\omega)/\omega$ of the $0.30$ wt \% PIB in PB test fluid.
The solid line is the fit of a four-mode Maxwell model to the data
and the dashed lines are the individual spectral contributions,
$(2\eta''/\omega)_{k}$. Error bars are given on the upper-left
corner of the graph. (b) Steady shear material functions,
$\eta(\dot{\gamma})$ and $\Psi_{1}(\dot{\gamma})$ of the test fluid.
The solid lines are fits of the four mode Giesekus model to the
data. Error bars are again shown in the upper left corner of the
plots. (c) Plot of the shear rate dependent Weissenberg number,
$We(\dot{\gamma}) \equiv  \lambda(\dot{\gamma})\langle \dot{\gamma}
\rangle$ as a function of shear rate for the $0.30$ wt\% PIB in PB
test fluid. An error bar which applies to the part of the curve with
shear rate larger than 0.25 s$^{-1}$ is also shown.}
\end{figure}

The dynamic shear flow material functions, $\eta'(\omega)$ and,
$2\eta''(\omega)/\omega$ of the 0.30 wt\% polyisobutylene (PIB) in
polybutene (PB) test fluid were measured in small-amplitude
oscillatory shear flow in a cone-and-plate configuration and are
shown in Figure 2(a). This linear viscoelastic data, in conjunction
with the steady data shown in Figure 2(b), was used to fit a
four-mode linear Maxwell model; the parameters of which define the
relaxation spectrum which we have presented elsewhere [14].
Each mode is seen to be dominant in a different
region of the spectrum; a given mode starts to shear-thin once a
critical frequency $\omega_{k} \sim \lambda_{k}^{-1}$ is exceeded.
In the case of, $\eta'$ a similar pattern can be observed for the
modal contributions, $\eta'_{k}$, originating from the high
molecular weight solute; however, these contributions are not shown
since the contribution of the non-frequency-thinning solvent,
$\eta'_{s}=\eta_{s}$, dominates over the other modes.

Steady shear flow material functions were measured
using the rheometer in the cone-and-plate
and in the parallel-plate configurations. Plots of
$\eta(\dot{\gamma})$ and $\Psi_{1}(\dot{\gamma})$ are shown in
Figure 2(b). The linear Maxwell fit indicated a decrease in
viscosity from $\eta_{0}$ = 110 Pa.s to 90 Pa.s at the maximum shear
rate, for which data was obtained, of 40 s$^{-1}$. Because of this
decrease of only 18\% in the viscosity over nearly three decades of
shear rate the test fluid closely approximated an elastic, constant
viscosity non-shear-thinning Boger fluid. The highest shear rate for
which material functions were measured was 40 s$^{-1}$; at shear
rates exceeding this value viscous heating would act to decrease
$\eta$ by more than 9\% and $\Psi_{1}$ by more than 18\%.
Substantial shear thinning is observed for the first-normal-stress
coefficient. The zero-shear-rate limit for $\Psi_{1}$ could not be
reached because of the sensitivity limits of the normal force
transducer. However, at 0.25 s$^{-1}$, the lowest shear rate for
which results reproducible to within 25\% could be obtained, the
data indicated a lower bound on the zero-shear-rate limit of
$\Psi_{1} > 200$ Pa.s$^{2}$. The nonlinear viscoelastic four-mode
Giesekus model is capable of capturing the shear thinning of the
material functions; values of $\alpha_{k}$, which control the shear
thinning behavior of the steady shear material functions, in
addition to the coefficients $\eta_{k}$ and $\lambda_{k}$ are
published elsewhere [14].

The stress-optical coefficients were determined using a Couette cell
apparatus. For a given shear rate, the stress components in the
fluid were determined from the previously measured viscometric
functions. The birefringence and extinction angle of the fluid
undergoing shearing were measured using the FIB system. The
refractive index tensor {\bf n} in the flow of a polymer blend is
related to the component polymer contributions to the stress tensor
as
$
{\bf n} = C_{\mathrm{PIB}}{\btau}_{\mathrm{PIB}} +
C_{\mathrm{PB}}\btau_{\mathrm{PB}}
$
where $C_{\mathrm{PIB}}$ is the stress-optical coefficient of the
polyisobutylene solute, and $C_{\mathrm{PB}}$ is the corresponding
coefficient for polybutene solvent. For all shear rates accessible
with the rheometer used, the polybutene solvent exhibited Newtonian
behavior. The difference in the normal components of the refractive
index tensor arose exclusively from the contribution of the
polyisobutylene solute component. From the experimental data it was
estimated that $C_{\mathrm{PB}} = 0.98 \times 10^{-9} \pm 0.07
\times 10^{-9}$ Pa$^{-1}$ and $C_{\mathrm{PIB}} = 1.48 \times
10^{-9} \pm 0.18 \times 10^{-9}$ Pa$^{-1}$. The effect of form
birefringence (see [14] for more details)on the measurement of $C_{\mathrm{PIB}}$ was minimal
since the isotropic refractive index of the polybutene solvent
matched that of the polyisobutylene solute, $n_{\mathrm{PB}} =
n_{\mathrm{PIB}} = 1.50$.

Use of the longest relaxation
time, $\lambda_{1}$, is appropriate for flows with low
characteristic shear rates, ${\dot{\gamma}}_{\mathrm{low}}$, such
that for the shorter relaxation times associated with the other
modes one obtains $\lambda_{k>1}\dot{\gamma} \ll 1$. However,
because of the shear thinning nature of the fluid, use of the
longest relaxation time will over-predict the characteristic
relaxation time for elevated flow rates.
We use the rate dependent
Weissenberg number
\begin{equation}
We \equiv \lambda(\dot{\gamma})\langle \dot{\gamma}
\rangle = {\psi_{1}(\dot{\gamma}) \over {2
\eta(\dot{\gamma})}}\langle \dot{\gamma}
\rangle.
\end{equation}
to characterize the importance of
elastic effects in our experiments. The effect of elasticity in the presence
of shear thinning taken into consideration
via the shear-rate dependent relaxation time $\lambda(\dot{\gamma})$.
A plot of $We(\dot{\gamma})$ is shown in Figure 2(c).
Predictions of $\Psi_{1}(\dot{\gamma})$ and
$\eta(\dot{\gamma})$ were obtained from the fitted Giesekus model.
Relaxation modes with time scales as great as 20 s were fit to the
shear rheology information; reproducible dynamic data were obtained
for frequencies as low as 0.1 s$^{-1}$. The plot shown is corroborated
by experimental viscometric material function data for $\dot{\gamma}
\geq 0.25$  s$^{-1}$. However, for smaller shear rates, the plot
represents an extrapolated prediction of the fitted Giesekus model.

The highest value of mean upstream shear rate attained in the 2:1
contraction experiments is
$\langle{\dot{\gamma}}_{\mathrm{Up}}\rangle = 4.9$ s$^{-1}$, which
corresponds to $We_{\mathrm{Up}} = 1.80$; this value is within the
range of the steady shear flow rheological data. The lowest value of
up-stream shear rate for which an experimental result (taken in the
32:1 contraction) is reported is
$\langle{\dot{\gamma}}\rangle_{\mathrm{Up}}  = 0.0021$ s$^{-1}$,
which corresponds to $We_{\mathrm{Up}} = 0.004$. The maximum
value of the Reynolds number based on downstream conditions for a
test run in this study is $7 \times 10^{-4}$ so that inertial
effects were negligible.

\section{Spatiotemporal transitions in the structure of flow states}

In general, a characteristic shear rate is
selected so that global transitions in the flow field associated
with elastic phenomena correspond with $We \sim O(1)$. However, in
flow regions with shear rates above the selected characteristic
value, local elastically induced flow phenomena may occur at
Weissenberg number much less than one. Consequently, a Weissenberg
number, $We_{\mathrm{Up}}$ , based on the upstream mean shear rate,
$\langle{\dot{\gamma}}_{\mathrm{Up}}\rangle = \langle v_{z} \rangle
/H$, is used to present our results.
Several considerations motivate this choice. Firstly,
$We_{\mathrm{Up}}$ can be determined solely from information on flow
conditions for a given test run and known shear rheological data.
Our experimental results indicate that onset of the instability is
associated with flow conditions (shear rate and streamline
curvature) upstream of the contraction plane. Furthermore, the
critical $We_{\mathrm{Up}}$  for onset of instability onset is
observed to decrease with increasing contraction ratio; in flows
through geometries with large contraction ratios, $H/h \geq 8$ for
instance, very low values of the critical $We_{\mathrm{Up}}$  viz,
$We_{\mathrm{Up,crit}} \ll 1$ are observed. This suggests that the
$We_{\mathrm{Up}}$  alone does not capture the physics controlling
onset of instability. It is more intuitive to select the
smallest length scale in the problem and consequently a Weissenberg
number based on downstream conditions to characterize the flow
patterns. In fact, we do interpret results in terms of both upstream
and downstream definitions and find that either by itself is
insufficient to characterize critical behavior. Consequently, we
choose $We_{\mathrm{Up}}$ as the parameter of study.

We begin with an examination of the
spatio-temporal structure of flow states and transitions
observed. Off-centerline scans in
space and time-series data acquired at a point are used to identify
the $We_{\mathrm{Up}}$ at onset of instability as well as associated
spatial wavenumbers and temporal frequencies.

\subsection{Velocity-field visualization}

\subsubsection{Flow transitions in the 8:1 contraction}

We first present data for the 8:1 contraction obtained from light
sheet visualization of the flow at different flow rates.
\begin{figure}
\includegraphics[width=14cm]{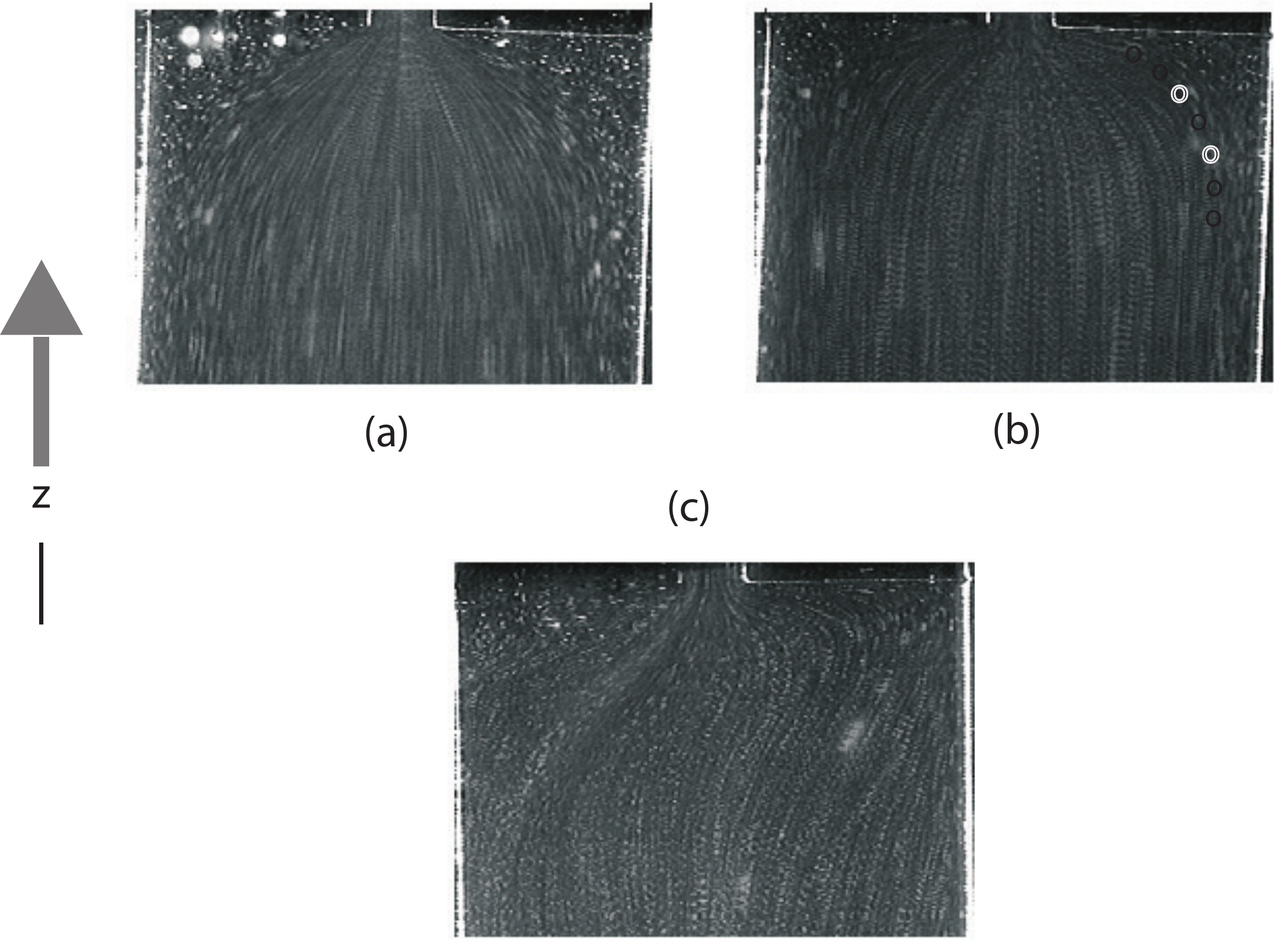}\
\caption{Side view of viscoelastic flow through the abrupt 8:1
planar contraction:(a) low flow rate with converging streamlines and
Moffat vortex in outer corner, $We_{\mathrm{Up}} = 0.052$ (number of
superimposed frames ($N_{\mathrm{frames}}$) = 15, time interval
between frames ($(\Delta t)_{\mathrm{frames}}$) = 0.33 s); (b)
reduction in size of corner vortex and flattening of streamlines
(shown by the sequence of circles following one streamline)
indicating development of diverging streamlines , $We_{\mathrm{Up}}
= 0.108$ ($N_{\mathrm{frames}}$ = 10, $(\Delta t)_{\mathrm{frames}}$
= 0.20 s). The  black circles
interspersed by white circles are a guide to the eye indicating the
onset of flattening.; (c) asymmetric, three-dimensional structure at high flow
rates, $We_{\mathrm{Up}} = 0.229$ ($N_{\mathrm{frames}}$ = 4,
$(\Delta t)_{\mathrm{frames}}$ = 0.12 s). Arrow marks the direction of the flow.}
\end{figure}
Snapshots of the $yz$-plane streakline velocity patterns at various
Weissenberg numbers are shown in Figure 3. For $We_{\mathrm{Up}} =
0.052$ as shown in Figure 3(a), the flow field of the test fluid
resembles the Newtonian flow profile. Specifically, the $yz$-plane
streakline image indicates that the streamlines converge smoothly
from the upstream channel into the downstream duct and that the flow
is steady and symmetric about the center-plane $(\upsilon = 0)$. The
$xz$-plane images give no indication of flow in the $x$-direction.
The resolution of the streakline images is insufficient to determine
quantitatively the reattachment length of the outer vortex ($L_{v}$)
in the upstream channel. However, the streakline image is consistent
with an eddy in the outer corner with reattachment length of
$L_{v}/H$ = 0.34. We note that this flow field has been predicted
numerically and analytically for Stokes flow ($Re = 0$) of a
Newtonian fluid [15,16].

As $We_{\mathrm{Up}}$ is increased,  the stream-lines still appear
symmetric about the center-plane; however, instead of uniformly
converging to the center-plane as the downstream slit is approached,
the streamlines indicate divergence especially in a region
approximately one half-height $H$ upstream of the contraction plane
($\zeta = 0$). This onset firsts manifests as a tendency of the
streamlines to flatten  rather than curve continuously. The distance
in the $z$-direction over which the streamlines exhibit the most
rapid continuous convergence from the up-stream channel to the
downstream slit moves closer to the contraction plane - cf. Figure
3(b)- than is seen for the lower $We_{\mathrm{Up}}$. An apparent
decrease in the reattachment length of the outer corner vortex is
associated with this streamline shift. More detailed observations
suggest that this apparent streamline pattern is  due to a
transition to steady, three-dimensional flow for values of
$We_{\mathrm{Up}}$ greater than approximately $0.124$. 

Above
$We_{\mathrm{Up}} = 0.171$, the flow becomes fully time-dependent as
well as three-dimensional.
Although the  flow is unsteady  for $We_{\mathrm{Up}}
> 0.171$, snapshots at certain instants in time
of the flow structure show features qualitatively similar to the
patterns seen for  $0.124 < We_{\mathrm{Up}} < 0.171$. Thus,
qualitatively, one can learn about the structure of the three
dimensional steady flow by looking at these snapshots. For instance, consider the
streakline image of the flow in the $yz$-plane after onset of the
spatial instability shown in Figure 3(c) for $We_{\mathrm{Up}} =
0.229$. On the left half of the image the streaklines continuously
converge from the upstream channel into the downstream slit. A
separated vortex in the outer left corner is evident; the
reattachment length of the vortex appears to be approximately
$L_{v}/H \approx 0.3$, similar to that for the Moffat eddy for
Newtonian flow. The in-plane flow speed
$(v^{2}_{x}+v^{2}_{y})^{1/2}$ in the region immediately adjacent to
the boundary of the left outer vortex is greater than in any other
part of the image; a very high velocity gradient exists in the
vicinity of the vortex boundary.
In the upstream region of the flow ($\zeta < -0.2$), the $v_{z}$
velocity component of the flow is uniform throughout the left half
of Figure 3(c), in contrast with the right side where the fast flow
is located adjacent to the outer vortex. In the left half of the
upstream region of the figure, streaklines initially approach the
outer left wall; they move in the positive y-direction, away from
the centerplane. When they near the outer left corner ($\zeta \geq
-0.2$), the streaklines abruptly change direction and travel in the
negative y-direction, following along the wall which defines the
contraction plane ($\zeta = 0$). No outer corner vortex could be
observed on the right side of the image. When the streamlines reach
the downstream slit, they change direction in order to enter the
slit. Near the right entry corner of the downstream slit, a small
but distinct vortex is observed. This lip vortex is separated from
any vortical flow which may be present in the outer right corner.
Although the most dramatic effect of the flow transition on the
velocity field is observed near the outer walls and the contraction
plane, the transition has a noticeable effect on the entire flow
field up to at least a distance of approximately 1.5$H$ before the
contraction plane. Specifically, the streaklines near the
center-plane appear sinuous: the mean direction of flow is in the
positive $z$-direction, and the streaklines have a sinusoidal shape
in the $y$-direction.

\begin{figure}
\includegraphics[width=16cm]{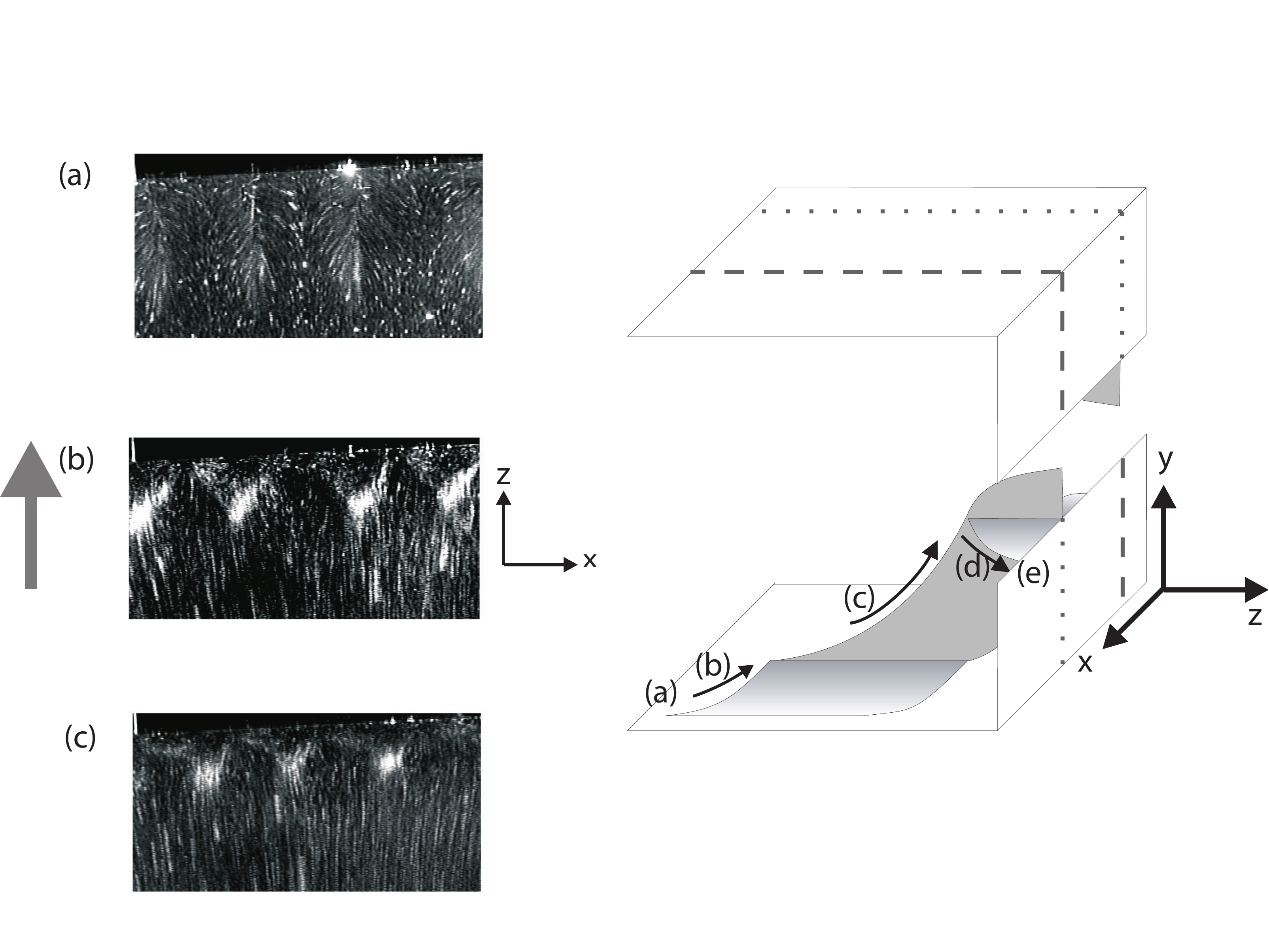}\
\caption{(A) Top view of the abrupt 8:1 planar contraction at high flow
rate, $We_{\mathrm{Up}}$ = 0.229 ($N_{\mathrm{frames}}$ = 15,
$(\Delta t)_{\mathrm{frames}}$ = 0.1 s for all images). Structure is
three-dimensional, slices in the $xz$ plane at different
$y$-elevations are shown: (a) flow near the outer wall ($\upsilon$ =
-7.5), fluid flows in the $x$-direction to feed regions of fast flow
visible as bright areas; (b) far from the contraction plane fluid
feeds the fast regions, near the contraction plane fluid spreads out
to assume a more uniform profile in the x-direction, $\upsilon$ =
-4.0; (c) at $\upsilon$ = -2.0, closer to the centerplane, the flow
is more uniform in the x-direction, within the triangular structures
fluid wells up from planes closer to the outer wall.
(B) Diagram representing the three-dimensional structure of the
flow after onset of the instability: (a) far upstream fluid near
wall fluid element travels in $z$-direction; (b) at distance order
$H$ before contraction plane element feeds into fast region; (c)
element in fast region flows toward downstream slit; (d) at distance
order $h$ before contraction plane fluid spreads out to achieve
$v_{z}$ more uniform along $x$; (e) fluid enters downstream slit.}
\end{figure}

\subsubsection*{Onset of temporal instability at high $
We_{\mathrm{Up}}$}

As $We_{\mathrm{Up}}$ is increased further beyond  a value equal to
0.171, onset of temporal instability is observed at a critical flow
rate. Thus the flow for $We_{\mathrm{Up}} = 0.229$ is time
dependent. 

Frozen images (taken at a particular
instant in time) of slices in the $xz$-plane located at positions ranging from
near the outer wall to near the center-plane for a Weissenberg
number of 0.229 are shown in Figure 4(A). These are discussed in order
of increasing $\upsilon$ position, starting from near the outer wall
and following the mean flow of fluid toward the downstream slit. At
a position $\upsilon  = -7.5$, streaks oriented in the $z$-direction
are visible. These streaks have a mean separation from each other in
the $x$-direction of about 1.5$H$; this separation defines the
wavelength of the three-dimensional flow. The fluid in the vicinity
of the streaks has a much faster $v_{z}$-component than fluid in the
intermediate regions. Moreover, fluid is observed to flow out of the
intermediate regions in the $x$-direction and feed into the streaks;
this accounts for the arced appearance of the streaklines on either
side of the fast streaks. In the slice at $\upsilon  = -4.0$, the
fast flow regions continue to be fed by the slow flow at distances
greater than 0.3$H$ upstream of the contraction plane. Near the
contraction plane the flow spreads out to assume a more uniform
profile of $v_{z}$ along the x-direction. Specifically, fluid
appears to travel along the bright boundaries which demarcate the
triangular structures between the contraction plane and the end of a
given streak. In the interior of these triangular structures, the
flow is directed primarily in the $y$ (out-of-plane) direction. An
image of a slice at $\upsilon =-2.0$ is shown in Figure 4(A-c). Here
the triangular structures are confined to a region nearer to the
contraction plane. Although the resolution of the streakline image
is limited, careful study of the source videotape indicated that
within the triangular structures there was flow in the
$x$-direction, directed away from the center of the structure. The
primary direction of flow within the triangular structures was still
in the $y$-direction, toward the center-plane. Throughout the image
(with the exception of the region located immediately before the
contraction plane, where the triangular structures are located) the
$v_{z}$ -component of the flow is more uniform along the
$x$-direction than the slices taken closer to the outer wall
(Figures 6(a) and 6(b)). Hence, close to the centerplane ($\upsilon
= 0$), the flow rearranges to adopt a more uniform velocity profile
along the x-direction before entering the downstream slit.

It was observed that regions of fast flow on a given side of the
center-plane correspond to regions of slow flow at the same
$x$-position on the other half of the center-plane indicating the
presence of bundles of counter-rotating vortex pairs having an
interlaced structure. This was confirmed by continuously moving the
$xz$ light sheet through the entire upstream height over a period of
45 s, much less than the time scale of the temporal oscillation.

From the light sheet slices in the $xz$- and $yz$-planes it is
possible to reconstruct a qualitative sketch of the
three-dimensional spatial structure of the flow valid as one extrapolates
to the critical flow rate at which temporality onsets. The history of a
fluid element traveling along a streamline which passes through the
fast region of flow is illustrated schematically in Figure 4(B). Fluid
near the wall of the upstream channel travels in the $z$-direction
of mean flow (a). When it reaches a distance of the order of the
upstream half-height, $H$, before the contraction plane, the fluid
begins to also travel in the $x$-direction, toward the fast region
(b). As it more closely approaches the fast region, the fluid begins
to travel in the y-direction, away from the outer wall and towards
the downstream slit (c). When the fluid approaches to within the
order of the downstream slit half-height, $h$, of the contraction
plane, it travels back in the positive $x$-direction, away from the
center of the region of fast flow (d). The magnitudes of $v_{x}$ and
$v_{y}$ are small when compared with that of $v_{z}$ when the fluid
reaches the contraction plane (e). The spreading out of the flow
near the contraction plane makes the $v_{z}$ component nearly
uniform along the $x$-direction near the contraction plane $\zeta
=0$.

The temporal structure of the instability is elucidated
by sets of $yz$ and $xz$ sheet images taken successively in time. An
image taken in the $yz$ plane of the flow through the 8:1
contraction is shown in Figure 5(a) and corresponds to
$We_{\mathrm{Up}} = 0.229$ at $t = 0$ s; the streaklines are
asymmetric with the region of fast flow on the right half of the
image. An image in the $xz$ plane of the flow taken with the light
sheet located near the outer wall, at $\upsilon  = -7.8$, is shown
in Figure 5(b). The $yz$ and $xz$ views were taken at different
absolute times, but at the same flow conditions. Dashed arrows on a
given image (e.g., $yz$ plane) indicate the position of the other
image (e.g., $xz$ plane). At $t = 105$ s, the streaklines (in Figure
5(c)) appear nearly symmetrical about the center-plane. The
corresponding $xz$ image - Figure 5(d) - shows that the vortices
have moved toward the center of the flow ( $\chi = 0$). The flow
field shown in Figure 5(e) at $t = 270$ s is again asymmetric;
however, the region of fast flow is now on the left half of the
image. The flow in Figure 5(f) shows that the difference between
Figures 5(e) and 5(a) is a result of the vortices having moved
farther toward the center of the flow. Specifically, the region of
fast flow in Figure 5(a) has been replaced by a region of slow flow
in Figure 5(e).

\begin{figure}
\includegraphics[width=15cm]{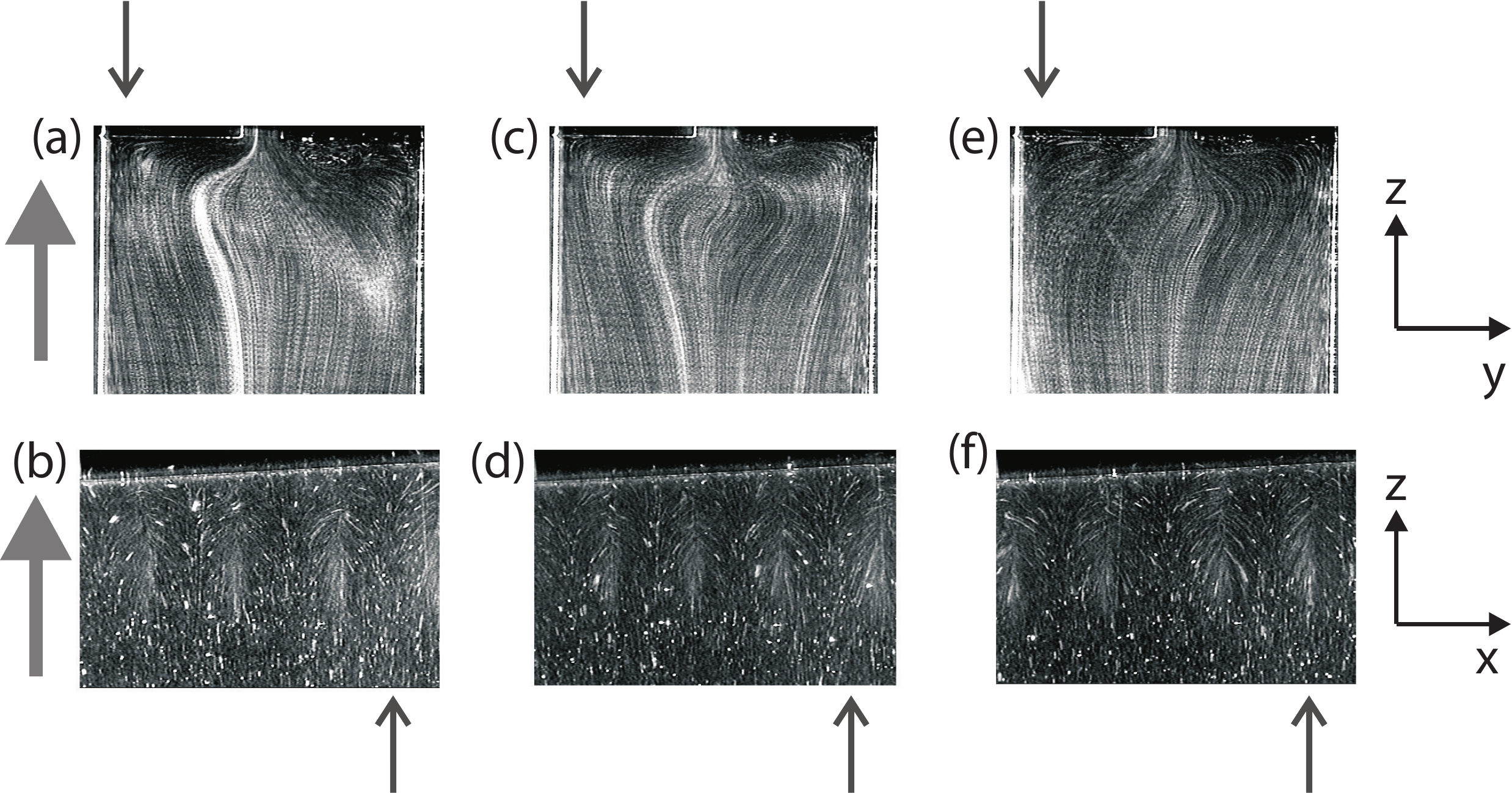}\
\caption{Successive images in time, $yz$- and $xz$-planes (fixed in
space) of the abrupt 8:1 planar contraction at high flow rate,
$We_{\mathrm{Up}}$ = 0.229 ($N_{\mathrm{frames}}$ = 15, $(\Delta
t)_{\mathrm{frames}}$= 0.1 s for $yz$-images, $N_{\mathrm{frames}}$
= 40, $(\Delta t)_{\mathrm{frames}}$ = 0.1 s for $xz$-images): (a,b)
views at t = 0 s, asymmetrical streaklines are shown in the
$yz$-image with region of fast flow on the right half of the image;
(c,d) views at t = 105 s, $xz$-image shows that vortex bundles have
moved toward the center of the flow ($\chi$ = 0), streaklines in the
$yz$-image appear nearly symmetrical; (e,f) views at t = 270 s, the
$yz$-image shows asymmetrical streaklines, now with the fast flow on
the left side of the image, $xz$-image shows that vortex bundles
have moved toward the center of the flow, the fast flow region in
(b) has been replaced by a slow flow region.}
\end{figure}

The flow closest to the walls bounding the $x$-directions ($\chi =
\pm 32$) could not be observed with light sheet slices in the $y$ or
the $xz$ plane; only positions in the range $-23 \leq \chi \leq 23$
were accessible. However, visual observations in conjunction with
the LDV measurements indicated that after onset of the temporal
instability the vortices in the flow are continuously born at the
walls bounding the x-direction ($\chi = \pm 32$) and move toward the
center ($\chi = 0$) of the flow. As a result, the mean spacing
between the vortex bundles decreases. Eventually a vortex bundle
must be destroyed to maintain an average spacing between the bundles
on the order of the upstream half-height. This occurs through one of
two mechanisms. Two neighboring streamline bundles near the center
of the flow may move closer to each other until they eventually
merge into one. Alternately, a vortex bundle near the center of the
flow may decrease in size and intensity as neighboring streamline
bundles on either side approach. Eventually the streamline bundle in
the middle disappears entirely; this process of absorption by
neighboring bundles, is shown in Figure 6 via successive $xz$-images
in time. A destruction mechanism which has elements of both the
merging and absorption processes, i.e. preferential absorption into
one of the neighboring vortex bundles, was also observed.
\begin{figure}
\includegraphics[width=10cm]{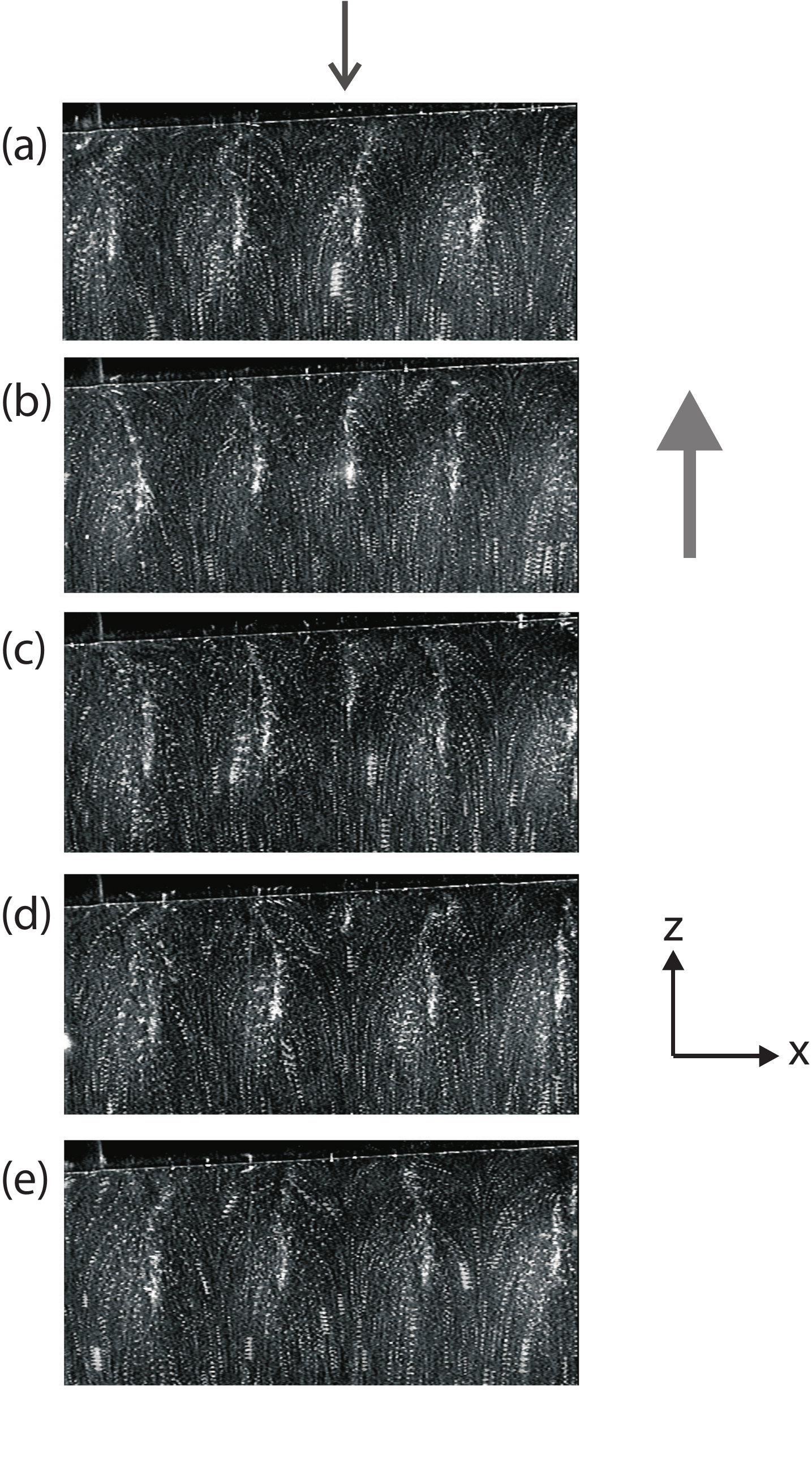}\
\caption{Successive streakline images in time, top view, of the
abrupt 8:1 planar contraction at high flow rate, $We_{\mathrm{Up}}$
= 0.229 ($N_{\mathrm{frames}}$ = 6): (a) t = 0 s, vortex bundles of
equal strength are evenly spaced $1.5H$ apart, note the arrow at top
which indicates the location of the center vortex bundle on which
attention is focused here; (b) view at t = 120 s, the vortex bundles
have moved toward the center of the flow ($\chi$= 0), resulting in
closer spacing; (c) at t = 180 s vortex bundles have moved yet
closer together and the center vortex bundle has weakened relative
to the neighboring bundles; (d) at t = 210 s the center vortex has
weakened further and is barely distinguishable; (e) at t = 240 s the
center bundle vortex has been completely absorbed into the
neighboring bundles.}
\end{figure}

\subsubsection{Flow transitions in the 2:1 and 32:1 contractions}

The flow transition sequence with increasing $We_{\mathrm{Up}}$  in
the 2:1 contraction was qualitatively similar to that observed in
the 8:1 contraction. However, the transitions occurred at higher
values of $We_{\mathrm{Up}}$ . The wavelength and the upstream
extent of the spatial instability were both of the order of the
upstream half-height, $H$, as found for the 8:1 contraction. There
were visual indications of time-dependent behavior of the vortex
bundles at elevated flow rates. The movement of the
vortices in the $xz$-plane was not as distinct as in the images
acquired with the 8:1 contraction flow.

In the 32:1 contraction, the low aspect ratio, $W/2H = 1$, of the
upstream channel affected the transition sequence and
spatio-temporal structure of the flow. Onset of three-dimensional
flow, specifically flow in the $x$-direction, could be detected via
light sheet visualization at the periphery of the observable region
encompassing $-23 \leq \chi \leq 23$. Additional detail on the
spatio-temporal structure of the flow after onset of the instability
is given in the sections to follow.

\subsection{Off-centerline velocity measurements of global flow
transitions}

Quantitative velocity field information was obtained to characterize
the spatial and temporal structure of flows after, and the class of
bifurcation at instability onset; the laser Doppler velocimetry
(LDV) technique was used. To parallel the qualitative observations
discussed above, results are given for transitions occurring at
increasing $We_{\mathrm{Up}}$. Off-centerline LDV data were taken
for the 2:1 and 32:1 contractions but not for the 8:1 contraction.

\subsubsection{\bf Transition to and evolution of three-dimensional
flow
in the 2:1 contraction}

The LDV system was operated with the frequency tracker in order to
characterize the wavelength of the three-dimensional flow field in
the $x$-direction after onset of the instability. Specifically, the
measuring volume was scanned through the region corresponding to
$-26 < \chi  < 0$, $\upsilon  =  -1.75$, $\zeta = -1.80$ with the
scan velocity $v_{x,{\mathrm scan}}$ held constant at 1.43 mm/s. The
velocity data measured as a function of time were then converted to
velocity as a function of spatial position. Data could not be
obtained for $\chi
> 0$ because the backscattered light used in the LDV had to travel
through too much fluid to yield a measurable signal.

As the volumetric flow rate was increased from an initially low
value, the structure of the flow field evolved spatially through a
sequence of transitions - first, onset of three-dimensional, steady flow, then
wavenumber doubling and subsequent appearance of multiple
harmonics.

\subsubsection*{Spatial profile of $v_{z}$ at intermediate
$We_{\mathrm{Up}}$}

At volumetric flow rates corresponding to $We_{\mathrm{Up}} \leq
0.37$, the profile of $v_{z}$ versus $x$ was uniform, and the flow
was two-dimensional. When the flow rate was increased to values
corresponding to $0.51 \leq We_{\mathrm{Up}} \leq   1.37$, the
profile of $v_{z}$ versus $x$ was no longer uniform, a transition to
a three-dimensional, steady flow had occurred.

The frequencies of the flow-structure information (0.1 Hz and 1.5
Hz) and instrument induced noise (about 14 Hz) were well separated,
and so a power spectrum (PS) of the data was used to identify
oscillations associated with the temporal structure of the
viscoelastic flow instability. The PS of the velocity versus
position data was calculated by using data in the range $- 26.1 <
\chi < -1.6$, for which the strongest signal was obtained. To obtain
a more accurate estimate of the PS, a set of eight velocity versus
position scans was made for a given flow rate. The Welch windowing
function was applied to each scan before calculation of the PS. The
eight power spectra were then averaged to generate a mean PS with a
standard deviation corresponding to 35 \% of a single PS.
\begin{figure}
\includegraphics[width=10cm]{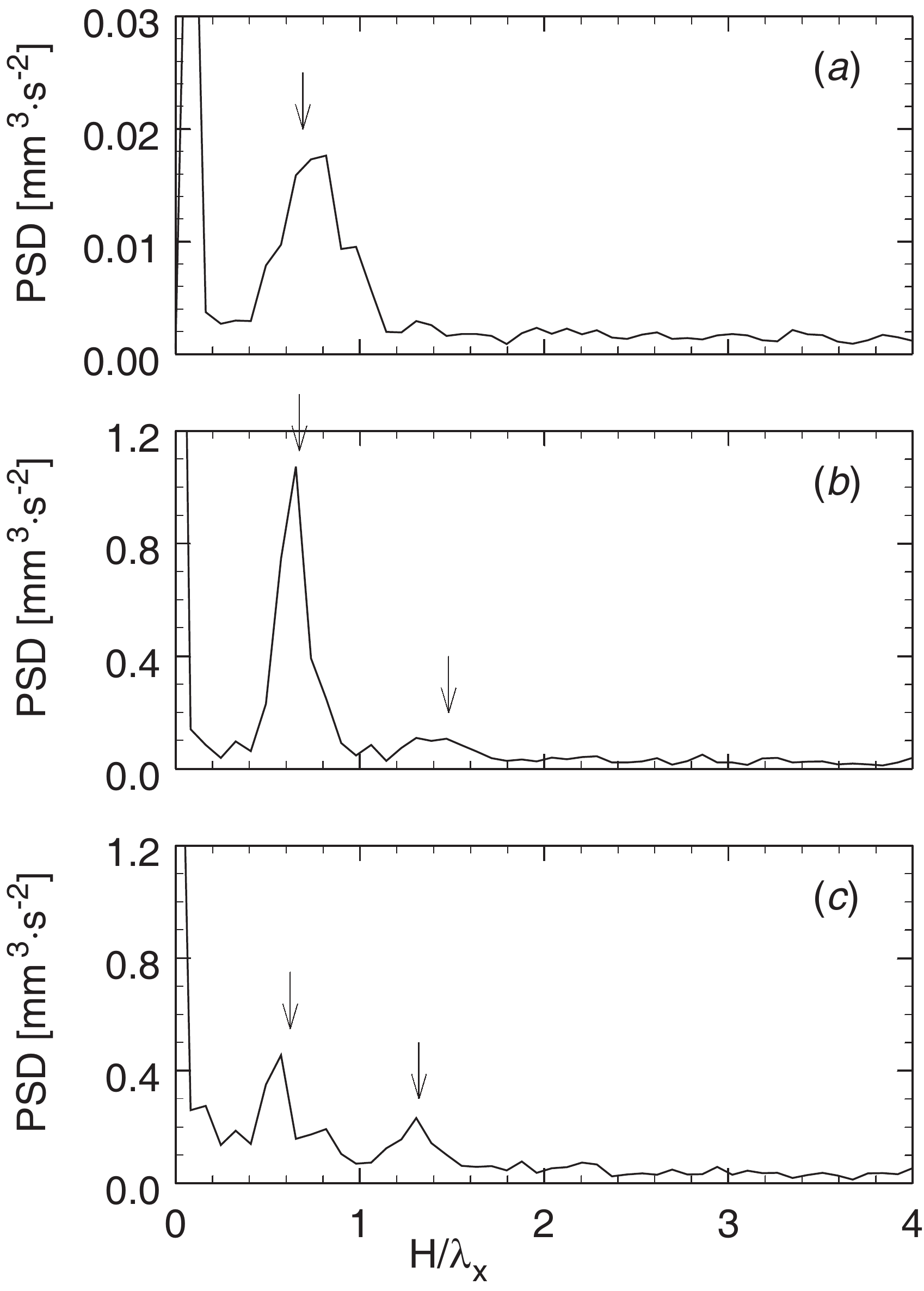}\
\caption{(a) Mean power spectrum (PS) of scan in the $x$-direction
($\upsilon$ = -1.75, $\zeta$ = -1.79) for flow through the 2:1
contraction with $We_{\mathrm{Up}}$ = 0.67. Mean dimensionless
wavenumber of primary peak is $\overline{(H/\lambda_{x})_{1}}=0.69$;
(b) PS of scan in $x$-direction ($\upsilon$ = -1.75, $\zeta$ =
-1.79), 2:1 contraction, $We_{\mathrm{Up}}$ = 1.53. Wavenumber
doubling behavior; primary peak wavenumber is
$\overline{(H/\lambda_{x})_{1}}=0.67$, secondary peak wavenumber is
$\overline{(H/\lambda_{x})_{1}}=1.48$; (c) PS of scan in
$x$-direction ($\upsilon$ = -1.74, $\zeta$ = - 1.80), 2:1
contraction, $We_{\mathrm{Up}}$ = 1.72. Primary peak wavenumber is
$\overline{(H/\lambda_{x})_{1}}=0.62$, secondary peak wavenumber is
$\overline{(H/\lambda_{x})_{2}}=1.32$. Higher order harmonics (in
addition to the primary and secondary peaks) are present. Note that
the mean dimensionless wavenumber of the peaks shown in (a - c) are
indicated by arrows.}
\end{figure}
A plot of the mean power spectral density (PSD) as a function of the
dimensionless wavenumber, $H/ \lambda_{x}$, where  $\lambda_{x}$ is
the wavelength in the $x$ direction is shown in Figure 7(a) for
$We_{\mathrm{Up}} = 0.67$. Wavenumber is used for the abscissa since
the PS is calculated for equal wavenumber intervals. The large peak
centered about $H/\lambda_{x} = 0.69$ corresponds to the
characteristic wavelength of the three-dimensional flow. The partial
peak at the extreme left of the plot is an artifact and is not
physically significant.

The power associated with the peak representing the spatial
oscillation in the PS was estimated by summing the power spectral
density (PSD) values of a continuous range of wavenumbers for which
the PSDs were greater than ten percent of the peak value. The
amplitude of a peak in the PS is defined as the square root of this
total power under the peak; the spatial oscillation shown in Figure
7(a) has an amplitude of 0.065 mm s$^{-1}$. Since the peak has a
finite breadth, an estimate of the wavenumber of the spatial
oscillation should consider the entire peak, not only the wavenumber
associated with the maximum power. Therefore, the wavenumbers were
weighted with their associated PSD values and summation performed
over the range. The sum was then normalized with the total power
over the range to determine the first moment corresponding to the
mean wavenumber. The characteristic dimensionless wavenumber of the
spatial oscillation calculated in this fashion  was $H/\lambda_{x} =
0.74$, which corresponds to a wavelength of $\lambda_{x} = 1.35H$.

\subsubsection*{Wavenumber doubling behaviour of spatial oscillation
in $v_{z}$ at high $We_{\mathrm{Up}}$}

For volumetric flow rates corresponding to $1.14 \leq
We_{\mathrm{Up}} \leq 1.53$, a secondary peak was observed at
approximately twice the spatial wavenumber of the primary peak. The
PS of the flow for $We_{\mathrm{Up}} = 1.53$ is shown in Figure
7(b). The primary peak has $H/\lambda_{x} = 0.67$; a broad
secondary peak with a dimensionless wavenumber of 1.48 is also
apparent.

\subsubsection*{Multiple harmonics in  spatial oscillation
in $v_{z}$ at high $We_{\mathrm{Up}}$}

The PS of a flow with $We_{\mathrm{Up}} = 1.72$ is shown in Figure
7(c). The secondary peak at $H/\lambda_{x} = 1.32$ has increased in
strength, whereas the primary peak at $H/\lambda_{x} = 0.62$ has
weakened. Additional peaks appear between the primary and secondary
peaks and below the primary peak. The PSD values for these
additional peaks are above the level of the broadband noise,
indicating that these peaks represent 
additional harmonics. However, the resolution of the PS (the
wavenumber interval) is limited by the finite distance in the
$x$-direction which can be probed, and the limited number of data
sets restricts the accuracy of the PSD estimate at a given
wavenumber. These considerations prevent specific identification of
the higher order harmonics.

\begin{figure}
\includegraphics[width=12cm]{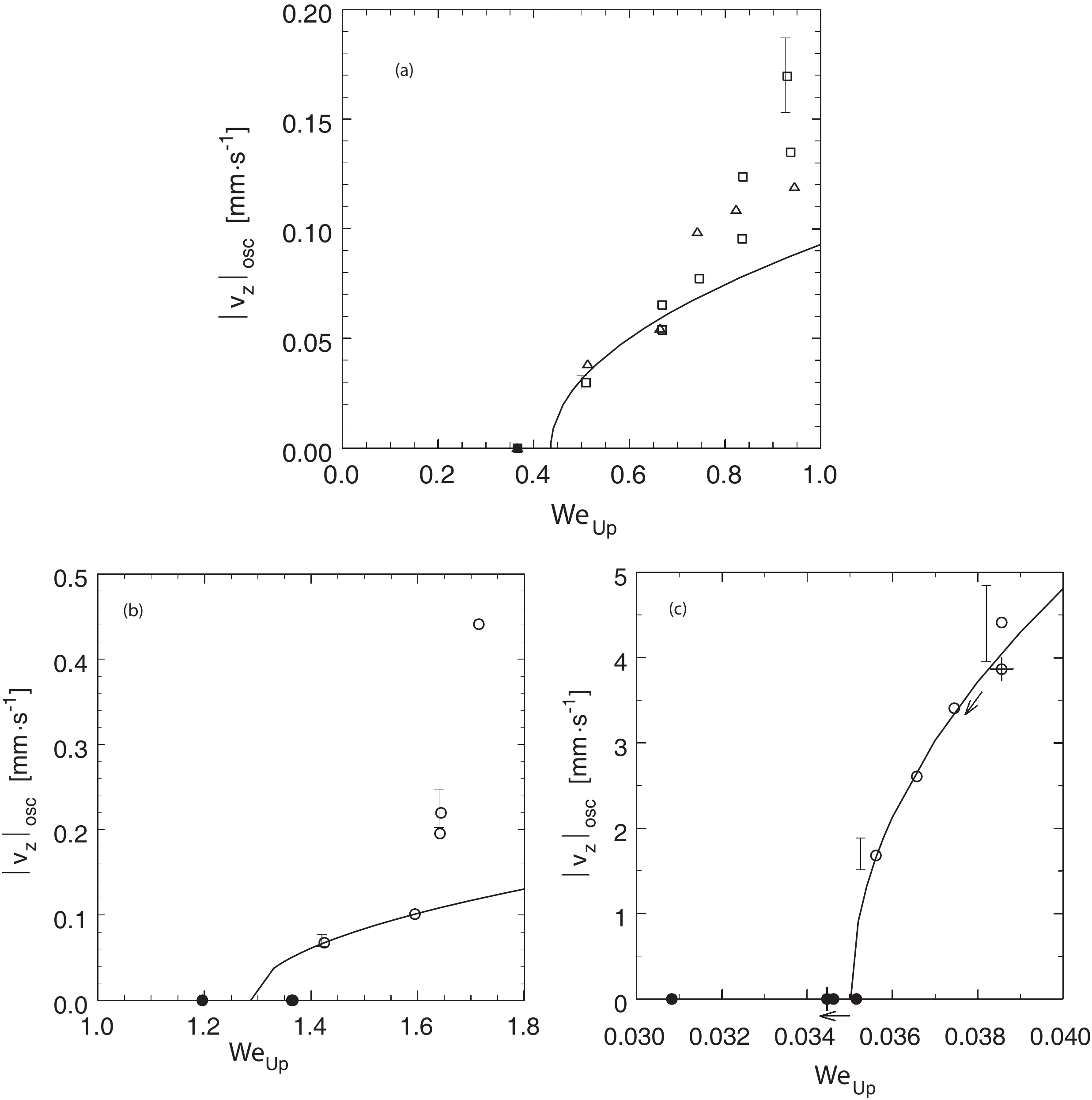}\
\caption{(a) Amplitude of spatial oscillation, $|v_{z}|_{osc}$, from
scans of $v_{z}$ vs. $\chi$ ($\upsilon$ = -1.75, $\zeta$ = - 1.80)
as a function of $We_{\mathrm{Up}}$ in 2:1 contraction flow . Data
($\Delta$) taken with successively  increasing flow rates; ($\Box$)
decreasing flow rates; note the absence of hysteresis. Solid symbols
indicate two-dimensional flow; hollow symbols denote
three-dimensional flow; (-), fit of equation to data near onset.
The order 10 \% error associated with the determination of the
amplitude of oscillation is indicated by the two representative
error bars. (b) Amplitude of oscillation, $|v_{z}|_{osc}$, for {\it temporal
instability} at the point $\chi$ = -20.0, $\upsilon$ = -1.75, $\zeta$
= -1.80 as a function of $We_{\mathrm{Up}}$ in 2:1 contraction flow.
Solid symbols indicate steady flow; hollow symbols denote
time-dependent flow; (-), square-root-scaling fit to data near
onset. The order 10\% error associated with determination of the
amplitude of oscillation is indicated by the two representative
error bars; (c) Amplitude of oscillation, $|v_{z}|_{osc}$ , for
time-dependent flow at the point $\chi$ =-21.0, $\upsilon$ =-1.50,
$\zeta$ =-1.50 as a function of $We_{\mathrm{Up}}$ in the 32:1
contraction. Solid symbols indicate steady flow; hollow symbols
denote time-dependent flow; (-), square-root-scaling fit to data
near onset. Note that the fit slightly under-predicts the value of
$We_{\mathrm{Up}}$ at onset of instability. (+) symbols and
associated arrows indicate that in one of the runs the volumetric
flow rate was decreased from $We_{\mathrm{Up}}$ = 0.0386 to
$We_{\mathrm{Up}}$ = 0.0345; hysteresis was not observed. The order
10\% error associated with determination of the amplitude of
oscillation is indicated by the two representative error bars.}
\end{figure}

A transition to time-dependent behavior occurs in the range $ 1.37
\leq We_{\mathrm{Up}} \leq 1.43$; the frequency associated with this
temporal oscillation is of order 0.006 Hz. In contrast, the output
from the tracker which contains the information on the spatial
period of the instability for the scan speed used corresponds to a
frequency between 0.1 and 1.5 Hz. Therefore, the time-dependent
behavior does not have a deleterious effect on the measurement of
the spatial form of the instability.

\subsubsection*{Classification of bifurcation to spatial
oscillation}

The amplitude of the spatial oscillation for flow
through the 2:1 contraction is plotted as a function of
$We_{\mathrm{Up}}$ in Figure 8(a). In order to check for the
occurrence of hysteresis, the volumetric rate was first set to a
value such that the flow was in the two-dimensional base state;
specifically $We_{\mathrm{Up}}\leq  0.37$. Measurements of the
velocity profile were then made at successively increasing flow
rates; the amplitude of a peak associated with the spatial
oscillation for a given $We_{\mathrm{Up}}$  is shown in the plot. In
a second set of runs, the flow rate was initially set to a value
such that the flow was in the three-dimensional and steady state.
Velocity profile measurements were then taken at successively
decreasing flow rates; the peak amplitudes for these measurements
are also shown. Since there is no significant difference between the
amplitudes of the sequences of measurements conducted for increasing
and for decreasing flow rates, no hysteresis is evident. The
standard deviation of the amplitude of oscillation of the eight
power spectra which were averaged together was of order 10\%; error
bars of 10\% are also indicated. The two-dimensional, steady base
flow is indicated by a filled symbol and three-dimensional flow by
hollow symbols.

Experimental observations indicate that in the absence of hystersis,
the bifurcation is likely to be a 
supercritical pitchfork bifurcation. For this type of bifurcation,
the amplitude of the spatial oscillation of the bifurcating
parameter, $v_{z}$, close to onset, is expected
from theoretical considerations [17-19], to scale with the control
parameter, $We_{\mathrm{Up}}$ as
\begin{equation}
|v_{z}|_{\mathrm{osc}} = C_{\mathrm{S2}}(We_{\mathrm{Up}} -
We_{\mathrm{Up,S2}})^{1/2},
\end{equation}
where $C_{\mathrm{S2}}$ is a constant. The letter in the subscript of parameters in
Equation (3) indicates the particular flow transition with which the
parameter is associated: S indicates transition from
two-dimensional, steady to three-dimensional, steady flow; T denotes
transition from steady to time-dependent flow. The number indicates
the contraction ratio, $H/h$, for which the parameter applies.
A similar equation also
arises  in instabilities of Newtonian fluids at high Reynolds
numbers [15,17].
Assuming that this expression indeed holds, we use equation (3) to fit
our experimental data near the value of $We_{\mathrm{Up}}$ for
onset. This procedure yielded $C_{\mathrm{S2}} = 0.124$ mm s$^{-1}$
and the critical value $We_{\mathrm{Up,S2}}= 0.44 \pm 0.07$.
The error bounds are derived from data points with
$We_{\mathrm{Up}}$ immediately greater than or less than
$We_{\mathrm{Up,S2}}$ for which stability of the base flow or
spatial oscillation was observed. For $We_{\mathrm{Up}}
> 0.74$, a departure from the square root scaling is clearly observed. We posit that
this behavior is associated with the onset of harmonics of the
fundamental wavenumber of the spatial oscillation.

\subsubsection{\bf Transition to Time-dependent flow in 2:1
contraction}

In order to study the temporal structure of the flow after
transition to time-dependent behavior, the measurement volume of the
LDV system was placed at the point $\chi$ =-20.0, $\upsilon =-1.75$,
and $\zeta$ =-1.80. This point was chosen since it was expected to
provide the clearest signal; specifically, in the spatial scan it
was within the region of maximum difference between the crest and
trough of the measured $v_{z}$. The flow rate was then set to a
value corresponding to a specific $We_{\mathrm{Up}}$, and the
velocity recorded as a function of time. The evolution in the
temporal structure of the flow is described here; the classification
of the bifurcation to time-dependent behavior is given at the end of
this section.

At volumetric flow rates corresponding to $We_{\mathrm{Up}} \leq
1.37$, no characteristic temporal oscillations were observed in the
power spectrum. Increasing the volumetric flow rate to a value
corresponding to $We_{\mathrm{Up}} \geq 1.43$ revealed oscillations
in the time-series data, which were characterized by calculating the
power spectrum in a similar manner as for the space data.
Overlapping data sets were extracted from the time series; the Welch
windowing function was applied to each set; individual PS were
calculated; and the results were averaged to compute a mean PS with
a lower standard deviation than the individual PS. The PS of at
least three time-series segments were averaged together, with each
segment consisting of approximately 900 data points representing
1500 s of flow. The mean frequency was calculated in a manner
analogous to the calculation of the mean wavelength for the spatial
instability. As an example, for $We_{\mathrm{Up}} = 1.72$, the
primary peak had a mean frequency of 0.0047 s$^{-1}$, which
corresponded to a period of 214 s.

Our experimental results indicate that the spatial oscillation of
the three-dimensional flow undergoes wavenumber doubling. For
$We_{\mathrm{Up}} \geq 1.80$ the power of the secondary peak in the
temporal PS exceeded that contained in the primary peak; an increase
in the amplitude of the secondary peak with $We_{\mathrm{Up}}$ and a
decrease for the primary peak also was noted in the PS of the
spatial oscillation discussed above.

We expect that the onset of periodic motion from the steady state is
via a super-critical Hopf bifurcation. Such a bifurcation could
either be subcritical (exhibiting hysteresis) or supercritical
(without hysteresis) [17]. 

For a super-critical Hopf-bifurcation, square-root
scaling of the amplitude near onset of temporal oscillation is
expected and we check our data to see if this is the case.
Let us first consider Figure 8(b) wherein the amplitude of the
temporal oscillation is plotted as a function of $We_{\mathrm{Up}}$.
Steady flow is indicated by the solid symbols; time-dependent flow,
by the hollow symbols. Noise in the data made it difficult to
characterize the amplitude of oscillation near the onset of the time
periodic instability. For the data points with representing
oscillation of finite amplitude, the error was of order 10\%, as
shown in the graph. A square root scaling of the form in Equation
(3) but now recognized as applying to temporal bifurcations was
applied to interpret the data. Unfortunately the measurement errors
inherent in the data do not allow us to either support or disprove
the scaling. One expects  a square root scaling
that is to be expected for a supercritical Hopf bifurcation, but the
data by itself supports a 1/3 power. It is possible however that
measurement errors and uncertainties might have resulted in this
result being a fit rather than the square root. Further measurements
are needed in the vicinity of the critical point. The fitted scaling
parameters for the temporal instability in the 2:1 contraction were
$C_{\mathrm{T2}} = 0.183$ mm s$^{-1}$ with $We_{\mathrm{Up,T2}} =
1.29$. Thus the square-root fit predicts a lower critical point for
the onset of instability than is observed.

\subsubsection{\bf Global flow transitions in the 32:1 contraction}

LDV measurements were restricted by the size of the side window of
the geometry to the region $-8 \leq \upsilon \leq 8$. Velocity
measurements near the wall of the upstream channel in the 32:1
contraction (at $\upsilon = \pm 32$) could not be obtained; hence,
in order to measure a large amplitude of oscillation associated with
a flow transition, the $v_{y}$-component of the flow was measured.
Spatial scans in the x-direction as well as time series information
at a given point were acquired with the LDV system operated in
frequency tracker mode. Flow phenomena which span the entire
$x$-dimension are considered here; phenomena localized near the wall
at $\chi$ = $\pm$ 32 are considered later. We discuss the spatial structure of
the instability in the 32:1 contraction first; then,
the temporal structure is addressed.

\subsubsection*{Spatial structure of flow after onset of
instability}

In contrast to flow in the 8:1 and 2:1 contractions, a transition
from the two-dimensional base flow to a three-dimensional, but
steady, spatial structure that encompassed the entire span of the
$x$-dimension was not observed. Rather, an immediate transition from
the base flow to a three-dimensional, time-dependent flow spanning
the $x$-dimension was observed.

LDV measurements were performed by moving the measuring volume in
the negative $x$-direction at a constant rate of $v_{x,scan} = 1.43$
mm s$^{-1}$. This scanning procedure did not lead to errors in
measuring length scales, since the voltage output of the tracker
during the spatial scans had a lowest characteristic frequency on
the order of 0.05 s$^{-1}$ compared to the characteristic frequency
of the temporal oscillation  of 0.003 s$^{-1}$. Scans performed at
different times for $- 32 \leq \chi \leq 0$, $\upsilon  = - 1.50$,
and $\zeta= -1.50$ for a flow with $We_{\mathrm{Up}} = 0.039$
indicated the formation of a wave originating at the bounding wall
at  $\chi =  -32$ and moving in the positive $x$-direction towards
the center of the flow. The results suggested that a temporal
oscillation was superimposed upon a steady, spatial oscillation. The
amplitude of the temporal oscillation seemed to be greatest near the
wall; this prevented accurate characterization of the wavelength,
amplitude, or onset $We_{\mathrm{Up}}$ of an underlying steady,
spatial oscillation. The amplitude of the temporal oscillation
was observed to decrease 
as the center of the flow was approached and the pattern of
the underlying spatial oscillation became distinct.

\subsubsection*{Characterization of onset and temporal structure of
flow transition}

To characterize the temporal structure of the instability in the
32:1 contraction, the measurement volume was placed at the point
$(\chi,\upsilon,\zeta) = ( -1.50, -21.0, -1.50)$, the flow rate was
set to a value corresponding to a specific $We_{\mathrm{Up}}$, and
the velocity recorded as a function of time. The evolution of the
temporal structure of the flow for successively greater Weissenberg
numbers was followed. Characterization of the dependence of the
magnitude of the oscillation on the Weissenberg number was used to
identify the critical value for onset of the instability.

For $We_{\mathrm{Up}} \leq 0.0352$ no time dependence of $v_{y}$ was
detected in the region scanned. When the volumetric rate was
increased to $We_{\mathrm{Up}} \geq 0.0356$, oscillations were
apparent in the time-series data. The PS was calculated and the mean
frequency of the oscillation peak was found to be 0.0029 s$^{-1}$.
The amplitude of oscillation is plotted as a function of
$We_{\mathrm{Up}}$ in Figure 8(c). For $We_{\mathrm{Up}}
> 0.0356$, the amplitude was observed to increase monotonically
until $We_{\mathrm{Up}} = 0.0424$; beyond this point the onset of
additional harmonics resulted in a decrease in the amplitude of the
primary peak. No hysteresis was noted when $We_{\mathrm{Up}}$
changed from $We_{\mathrm{Up}} = 0.0386$ to $We_{\mathrm{Up}} =
0.0345$, which was just below the critical point.

The amplitude as a function of $We_{\mathrm{Up}}$ near the critical
value was interpreted using 
\begin{equation}
|v_{z}|_{\mathrm{osc}} = C_{\mathrm{T32}}(We_{\mathrm{Up}} -
We_{\mathrm{Up,T32}})^{1/2},
\end{equation}
with $C_{\mathrm{T32}} = 68.2$ mm s$^{-1}$ and $We_{\mathrm{Up,T32}}
= 0.0350$ being the best-fit parameters. For $We_{\mathrm{Up}} =
0.0352$ the flow is steady; hence, equation (4) slightly under
predicts the critical $We_{\mathrm{Up}}$ for onset of instability.
This discrepancy is most likely to be  a consequence of error in the
measurement. Notice however that unlike in the Figure 8(b), the
agreement between the square-root scaling anticipated and that
observed is very good indicating that the bifurcation corresponds to
a supercritical Hopf bifurcation. Thus in summary, the temporal
oscillations, the lack of hysteretic behavior and the expectation of
a supercritical Hopf bifurcation motivate the square root scaling of amplitude
at onset.

For high volumetric rates corresponding to $We_{\mathrm{Up}} \geq
0.0424$, pronounced secondary harmonic peaks are visible in the PS.
The characteristic frequencies of the secondary peaks adjacent to
the primary peak were consistent with doubled and halved harmonics
of the fundamental frequency. However, the signal-to-noise ratio of
the velocity data was too low to allow more detailed classification
of the time-periodic flow in the nonlinear regime.

\subsection{Local evolution of flow near the bounding wall at $\chi
=\pm 32$ in the 32:1 contraction}

We also observed a viscoelastic flow transition localized near the
wall bounding the $x$-dimension for flow through the 32:1
contraction. The spatial structure of this transition was
characterized by making LDV measurements with the frequency tracker.
The measuring volume was scanned in the x-direction at a constant
rate of $v_{x,scan} = 1.43$ mm/s through the region of space
corresponding to $-32 \leq \chi \leq  0$, $\zeta = -1.50$ for
$\upsilon$ = -1.50 and +1.50. The acquired set of velocity versus
time data was then converted to velocity versus spatial position
data. Confidence in the profile was improved by reproducing the
measurements; a set of eight scans was used to develop a given
profile.

\begin{figure}
\includegraphics[width=12cm]{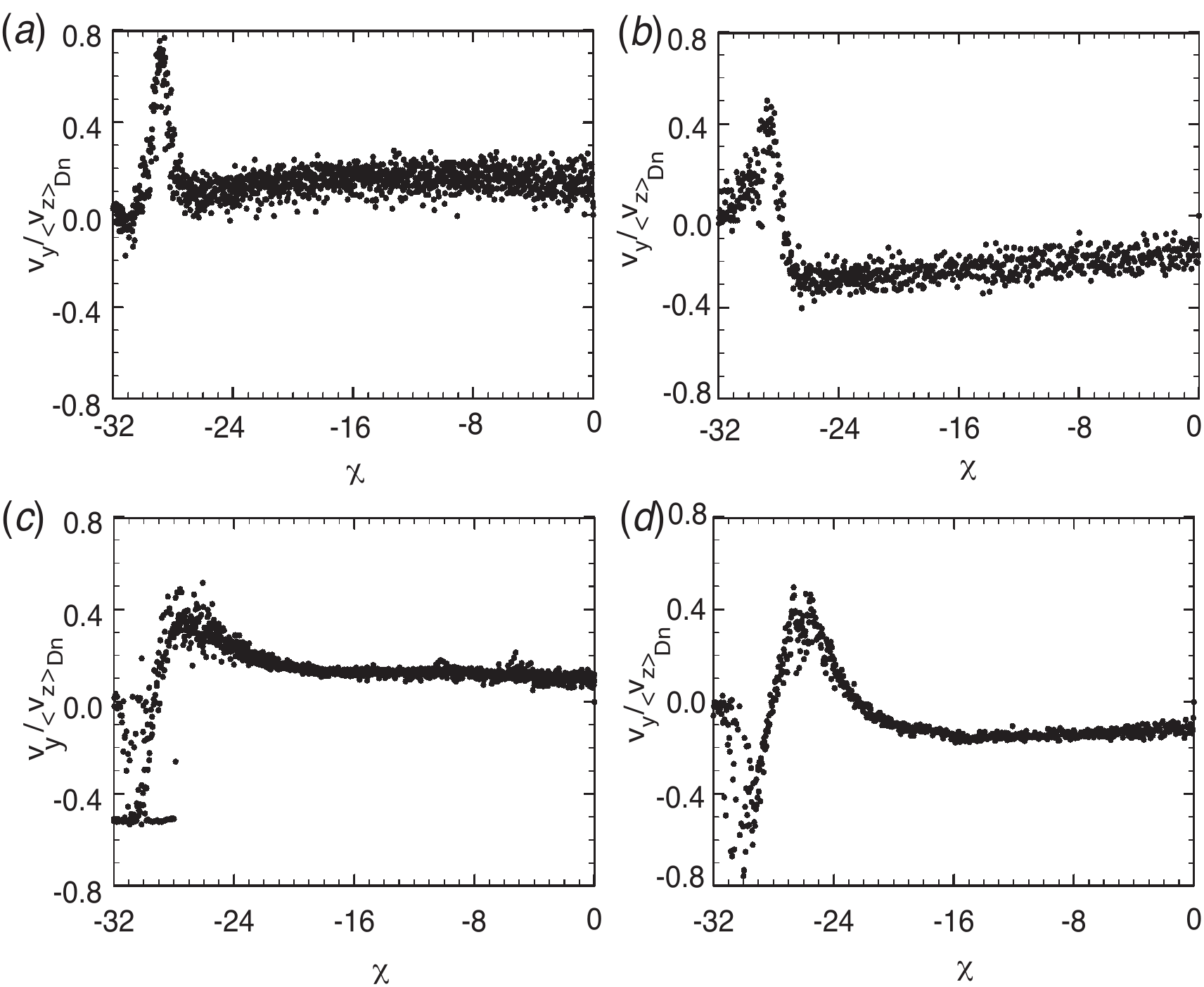}
\caption{Dimensionless $v_{y}$ vs $\chi$ plots for flow through the
32:1 contraction (a) $We_{\mathrm{Up}} = 0.006$,  $\upsilon  =
-1.50$, (b) $We_{\mathrm{Up}} = 0.006$,  $\upsilon  = +1.50$, (c)
$We_{\mathrm{Up}} = 0.031$,  $\upsilon  = -1.50$, and (d)
$We_{\mathrm{Up}} = 0.031$,  $\upsilon  = +1.50$.}
\end{figure}

The profiles for $We_{\mathrm{Up}} = 0.006$ are shown in Figures
9(a) and 9(b). In the middle section, $- 24 \leq \chi \leq 0$, the
flow appears uniform to within the accuracy of the measurement. The
velocity is positive for $\upsilon  =  -1.50$ and the fluid flows up
toward the center-plane; for $\upsilon  = +1.50$, the velocity is
negative. The signal-to-noise ratio is low in these measurements
because of the slow flow rate. Near the wall, $-32 \leq \chi \leq
-24$ the flow is nonuniform in the $x$-direction. In particular, the
flows in the middle section ($-24 \leq
\chi \leq 0$) on either side of the center-plane are in opposite
directions; however, the peak of the spike, located at $\chi =- 29$,
is positive for both $\upsilon = -1.50$ and $\upsilon = +1.50$. The
spike is neither a measurement artifact nor a transient resulting
from finite response time of the tracker electronics.

The velocity profile is illustrated in Figures 9(c) and 9(d) along
the $x$-direction for $We_{\mathrm{Up}} = 0.031$. The flow is still
steady; the onset of time-dependent behavior occurred at
$We_{\mathrm{Up,T32}} = 0.036$. The velocity profile near the center
of the flow ($-20 \leq \chi \leq 0$) is uniform and in opposite
directions for $\upsilon = - 1.50$ and  $\upsilon = +1.50$. The
nonuniform region of the profile has expanded to fill the range $-32
\leq \chi \leq - 20$ and both positive and negative values of
$v_{y}$ are seen. Specifically, for the range $ -32\leq \chi < -28$
the velocity is negative on both sides of the center-plane, while
for $-28 < \chi \leq -24$ the velocity is positive on both sides.

The velocity profiles observed are consistent with the flow near the
$\chi$ = -32 wall having a vortex structure in the $xy$-plane. At
the lower flow rate, $We_{\mathrm{Up}} = 0.006$, the elliptic point
of the vortex is located close to the wall, in the range  $-32 \leq
\chi \leq -30.5$. At higher flow rates, the vortex increases in
size, with the elliptic point located within the region $-29.5 \leq
\chi \leq -28$. This vortex growth is a non-Newtonian phenomenon. At
the low flow rate, $We_{\mathrm{Up}} = 0.006$, the phenomenon is
probably purely local, in extent as well as origin; the extent of
the vortex is on the order of the downstream length scale, $h$, not
the upstream length scale, $H = 32h$.

In the vicinity of the downstream slit, the fluid near the wall
experiences a higher shear rate than fluid at other locations in the
$x$-direction with the same $y$ and $z$ values because of the
additional no-slip boundary. This higher shear rate results in onset
of an instability which is localized near the wall. At higher flow
rates, we expect the global flow to interact with the local
instability. When wall effects are negligible and two-dimensional
flow is closely approximated throughout the geometry, the length
scale of the spatial instability is set by the upstream half-height,
$H$. However, for the 32:1 contraction ratio, the width of the
geometry is only $W = 2H$. Therefore, the increased extent of the
vortex at $We_{\mathrm{Up}} = 0.031$, may result from expansion of
the local vortex near the wall, or could be a manifestation of the
global spatial instability observed for the 2:1 and 8:1 contractions
additionally modified by the presence of the bounding walls located
at $x = \pm 32h = \pm H$.

In flows through the 2:1 and 8:1 contraction geometries, the
upstream aspect ratios are sufficiently high ($W/2H =$ 16 and 4,
respectively), that the presence of a bounding wall at the extremes
of the $x$-direction can be treated as a local imperfection to the
two-dimensional base flow and to flow transitions at elevated
Weissenberg number. For the 32:1 contraction the upstream
aspect ratio is unity; consequently, the walls bounding the
$x$-direction affect the flow structure and the transition sequence
throughout the upstream channel. In the 2:1 and 8:1 contractions,
the local instability near the bounding walls at $\chi  = 32$ noted
in the 32:1 contraction is also anticipated to occur. The global time-dependent
flow may be influenced by the local flow transition which acts as a
perturbation that sets the direction of movement of the vortices,
from the bounding walls toward the center of the flow, $\chi = 0$.

\subsection{Quantitative characterization of centerline velocity
profile evolution}

The flow transition to three-dimensional, steady flow described
earlier appears to be associated with changes in the centerline
profile. This phenomenon was studied in detail by operating the LDV
system using burst analysis to measure the centerline velocity
profile at several flow rates for the 2:1, 8:1, and 32:1
contractions. Upstream of the contraction plane, a decrease in the
centerline velocity below the value for the fully developed upstream
channel flow was noted at elevated $We_{\mathrm{Up}}$. Close to the
contraction plane, in the entry flow region, an increase in the
local elongational strain rate was observed.

\subsubsection{Centerline velocity profile evolution in the 2:1
contraction}

The magnitude of the $v_{z}$ velocity component along the centerline
in the middle of the flow was measured for the 2:1 contraction;
specifically, scans were performed over the range $(\chi =
0,\upsilon  = 0)$, and $-40 \leq \zeta \leq 12$. The velocity data
as a function of position, $v_{z}(z)$  was fit to a cubic spline.
From this, the strain-rate profile as a
function of position, $\dot{\epsilon}  =
\partial{v_{z}}/\partial{z}$ was obtained. The maximum dimensionless strain rate,
$\dot{\epsilon}_{\mathrm{max}}( h/{\langle v_{z}
\rangle}_{\mathrm{Dn}})$, was used to quantify the sharpness of the
peak in the centerline strain-rate profile.

\begin{figure}
\includegraphics[width=15cm]{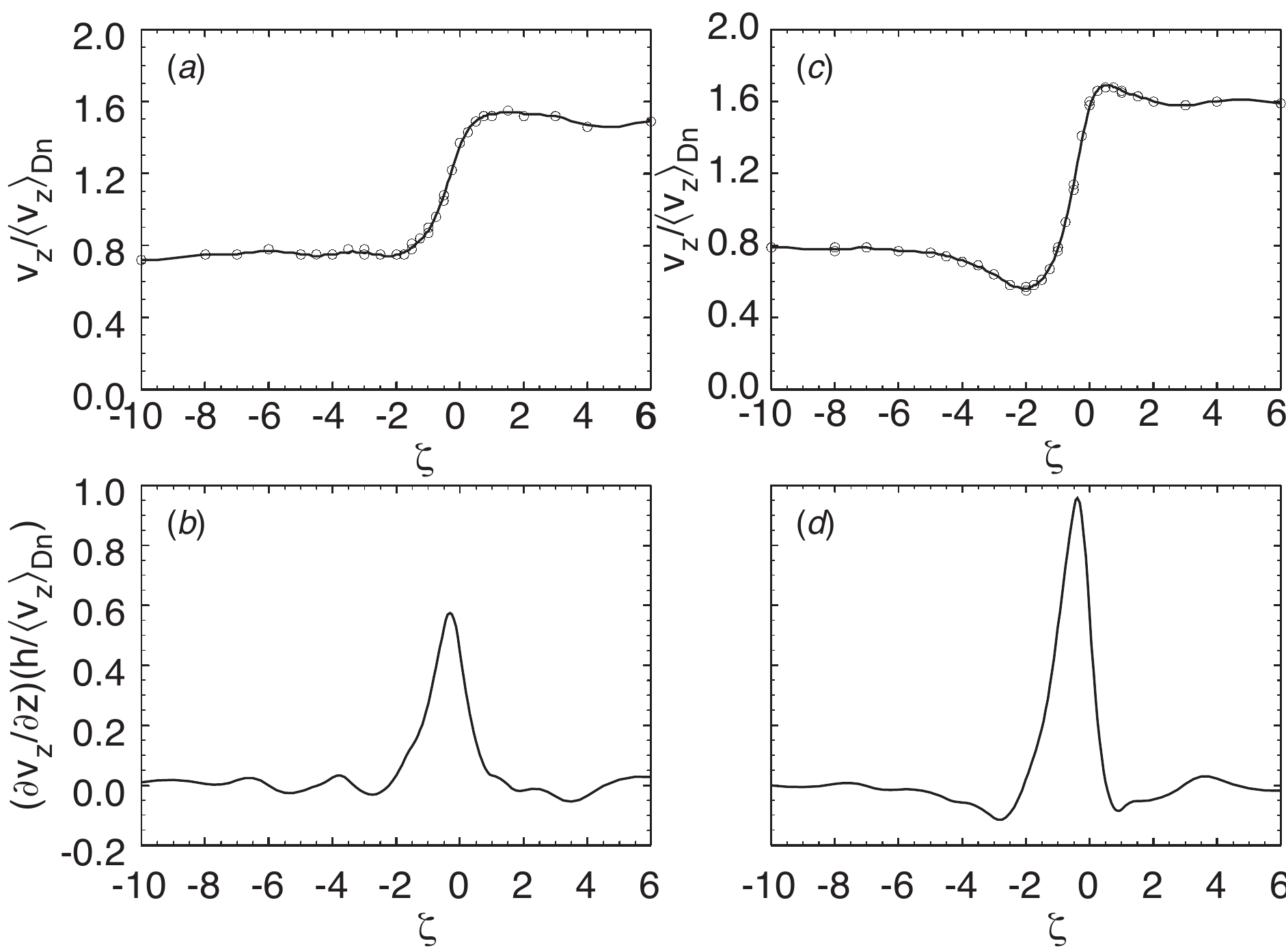}\
\caption{Centerline ($\upsilon$ = 0) velocity and elongational
strain-rate profiles for 2:1 contraction flow: (a) Dimensionless
$v_{z}$  vs. $\zeta$,  $We_{\mathrm{Up}}$ = 0.37; (b) Dimensionless
$\dot{\epsilon}$ vs. $\zeta$, $We_{\mathrm{Up}}$ = 0.37; (c)
Dimensionless $v_{z}$ vs. $\zeta$, $We_{\mathrm{Up}}$ = 1.20; (d)
Dimensionless $\dot{\epsilon}$ vs. $\zeta$, $We_{\mathrm{Up}}$ =
1.20. $(o)$ velocity data; (-) cubic spline fit.}
\end{figure}

Results for $We_{\mathrm{Up}} = 0.37$ are shown in Figure 10(a). Far
upstream of the contraction plane ($\zeta < -8$), the profile in the
$y$-direction is parabolic.
The centerline velocity made dimensionless with the
downstream mean velocity is approximately $v_{z}/{\langle v_{z}
\rangle}_{\mathrm{Dn}} \approx {3 \over 2}h/H$. When a fluid element
approaches to within two down-stream half-heights of the contraction
plane ($\zeta  > -2$) it suddenly accelerates to the downstream
fully developed centerline velocity, $v_{z} \approx {3 \over
2}{\langle v_{z} \rangle}_{\mathrm{Dn}}$. The dimensionless strain
rate $(\partial{v_{z}}/\partial{z})(h/\langle v_{z}
\rangle_{\mathrm{Dn}})$ is shown in Figure 10(b) as a function of
axial position. The acceleration of the fluid occurs over the range
$-2 \leq \zeta \leq 1.5$. The maximum dimensionless strain rate of
0.57 is observed at a position $\zeta= - 0.31$, immediately
up-stream of the contraction plane.
At a higher flow rate corresponding to $We_{\mathrm{Up}} = 1.20$,
diverging flow is evident along the centerline. As shown in Figure
10(c), the fluid initially {\em decelerates} from the fully
developed upstream velocity and then {\em accelerates} near the
contraction plane. Another contrast with the centerline velocity
profile for low $We_{\mathrm{Up}}$ is that a pronounced velocity
overshoot is observed immediately downstream of the contraction
plane. The centerline strain-rate profile for $We_{\mathrm{Up}} =
1.20$ is shown in Figure 10(d). Diverging flow is evident as a
region of negative strain rate for $- 6 \leq \zeta \leq  -2$; the
minimum centerline velocity occurs at $\zeta = -2.1$, and the
velocity overshoot appears as a region of negative strain rate
located downstream of the con-traction plane which extends over the
range $0.5 \leq \zeta \leq 3$.

The maximum dimensionless strain rate achieved for high
$We_{\mathrm{Up}}$ flow is substantially greater than for low
$We_{\mathrm{Up}}$. This effect is attributable to two factors.
Firstly, the difference between the minimum upstream and maximum
downstream dimensionless velocity is greater for the higher value of
$We_{\mathrm{Up}}$ because of the diverging flow and velocity
overshoot. Secondly, the region where the fluid accelerates from the
upstream into the downstream slit decreases slightly in extent, from
$-2 \leq \zeta \leq 1.5$ for $We_{\mathrm{Up}} = 0.37$ to $-2 \leq
\zeta \leq 0.5$ for $We_{\mathrm{Up}} = 1.20$. Note that this flow
rearrangement is related to the elastic nature of the flow since the
Reynolds number is low.

\subsubsection{Centerline Velocity Profile Evolution in 8:1 and 32:1
contractions}

The profiles of the centerline velocity and the corresponding
centerline strain-rate were also studied for flow through the 8:1
and 32:1 contractions. At $We_{\mathrm{Up}} = 0.070$, the
strain-rate profile for the 8:1 contraction appears qualitatively
similar to that for the 2:1 contraction; however, in the 8:1
contraction the axial region over which the flow accelerates from
the upstream into the down-stream slit is larger. Additionally, in
contrast with the 2:1 contraction flow, the strain-rate profile is
composed of two distinct regions, a high strain rate region and a
low strain rate upstream tail. These two regions become more
pronounced as the contraction ratio is increased to 32:1. The
precise distance that this tail extended upstream could not be
determined for the 32:1 con-traction due to restrictions in the LDV
measurements. For example, in the 32:1 contraction at a Weissenberg
number of 0.038, the region of greatest strain rate is restricted to
$-2 \leq \zeta \leq 0.25$. As the Weissenberg number increases, the
distinction between the high strain-rate region and low strain-rate
tail becomes clearer.

\subsubsection{ Evolution of Centerline Velocity Profile}

The centerline profiles for the flows through the different
contraction ratios exhibit similar qualitative
features. At low $We_{\mathrm{Up}}$ , there are two distinct regions
of positive strain rate through which a fluid element on the
centerline passes when accelerating from the slit upstream of the
contraction plane to the slit downstream of the contraction plane. A
low strain-rate upstream tail is evident in the profiles for the
32:1 and 8:1 contractions. The upstream extent of the tail appears
to be set by the upstream half-height, $\zeta \sim H/h$ (although
the restriction on LDV measurement to $\zeta \geq -40$ did not allow
identification of the precise extent in the 32:1 contraction). At
the location $\zeta \approx  -3$, upstream of the contraction plane,
the tail region adjoins a high strain-rate peak. The location of the
transition between the low strain-rate tail and high strain-rate
peak appears to be set by the downstream half-height, $h$.

The existence of two distinct regions, one a long, low strain-rate
upstream tail, the other a high strain-rate peak can be explained as
follows. On the centerline and near the contraction plane, the flow
field is governed primarily by the downstream slit; the region of
high strain rate, the extent of which is set by the downstream
half-height, is characteristic of this entry flow field. Stated
another way, the flow near the slit does not see the upstream
boundary conditions. Following this reasoning, a limiting profile is
expected near the contraction plane for the case of an infinite
contraction ratio (sink flow). However, for a flow with a finite
contraction ratio, the upstream half-height, $H$, must play some
role in determining the strain-rate profile. Specifically, the net
Hencky strain experienced by a fluid element traveling along the
centerline from the up-stream to the downstream regions is set by
the ratio of centerline velocities of the upstream and downstream
fully developed flows or
\begin{equation}
\epsilon_{\mathrm{Hencky}} = \ln{({{v_{z}(\upsilon =
0)_{\mathrm{Dn}}} \over v_{z}(\upsilon = 0)_{\mathrm{Up}}})}.
\end{equation}
At a sufficient distance upstream of the contraction plane, the
upstream boundary conditions set the profile for the low strain-rate
up-stream tail, the spatial extent of which scales with the upstream
half-height, $H$. A distinct transition between the upstream tail
and downstream peak is not noted for the case of the 2:1
contraction, probably because the upstream and downstream
half-heights are sufficiently similar that the two regions overlap.

A common set of flow phenomena also is noted for the flows through
the different contractions at elevated $We_{\mathrm{Up}}$.
Specifically, in contrast with the flows for low $We_{\mathrm{Up}}$,
there is an increase in the maximum strain rate attained and a
velocity overshoot is observed immediately downstream of the
contraction plane. The spatial extent of the velocity overshoot in
the downstream region is governed by the downstream half-height,
$h$. This scaling is probably a result of the strain rate in the
vicinity of the downstream slit being much higher than that in the
low strain-rate upstream tail; the fluid only remembers flow
conditions near the downstream slit.

\begin{figure}
\includegraphics[width=15cm]{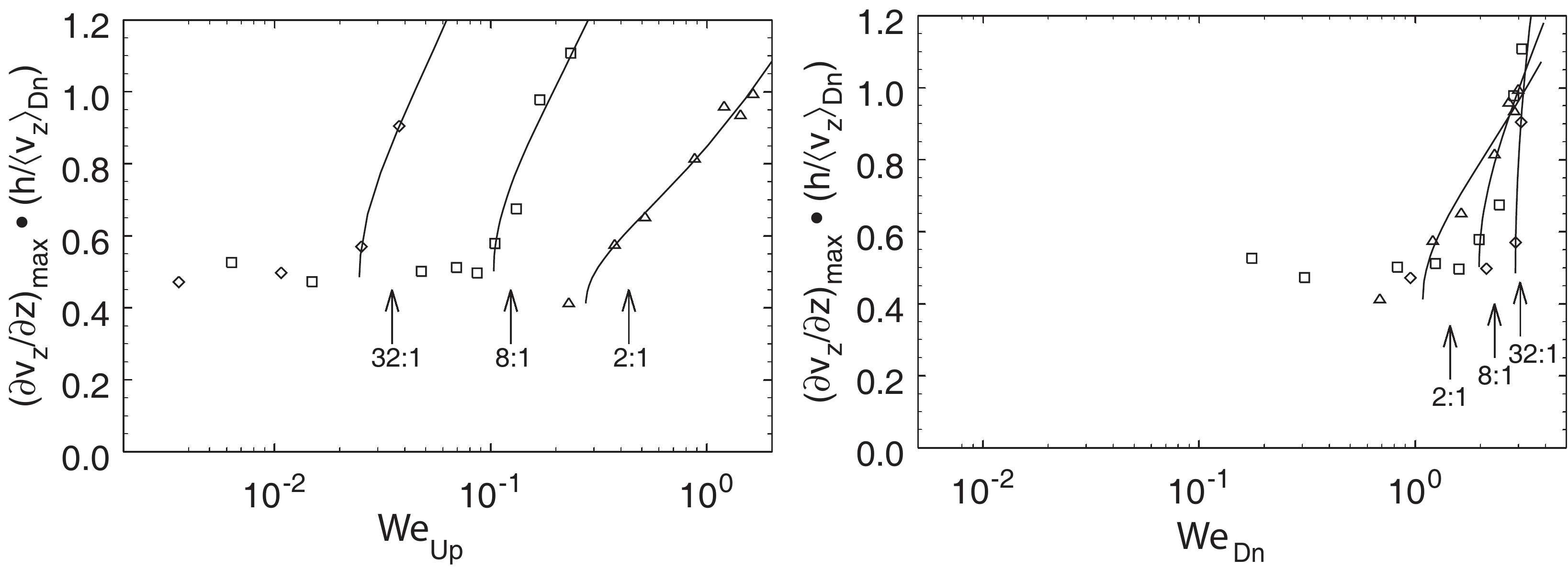}
\caption{Maximum centerline dimensionless strain rate vs.
Weissenberg number: ($\Delta$) 2:1 contraction; ($\Box$) 8:1
contraction; ($\lozenge$) 32:1 contraction. The lines are
square-root fits to the data, the lower terminus of the line is the
predicted onset point for the increase in dimensionless strain rate
with Weissenberg number; the arrows indicate the value for which
transition from global, two-dimensional base flow to
three-dimensional flow occurs. (a) domain is Weissenberg number
defined in terms of upstream flow parameters, $We_{\mathrm{Up}}$;
(b) domain is Weissenberg number defined in terms of downstream flow
parameters, $We_{\mathrm{Dn}}$.}
\end{figure}

At elevated flow rates, diverging flow was noted upstream of the
contraction plane for the 2:1 and 8:1 contractions. The point of
minimum centerline velocity occurred for the 2:1 and 8:1
contractions at the boundary between the diverging
($\partial{v_{z}}/\partial{z} < 0$) and accelerating
($\partial{v_{z}}/\partial{z} > 0$) flow regimes. The location of
this minimum velocity point appeared to scale with the upstream
half-height, $H$, moving further upstream with increasing $H$.
Specifically, for the 2:1 contraction, the minimum was at $\zeta$ =
-2.1; for the 8:1 contraction, at $\zeta$ = -5.5. No diverging flow
was seen for the 32:1 contraction flow; however, since the farthest
point upstream which could be probed was located at $\zeta$ = -40,
this observation is consistent with a scaling with the upstream
half-height, $H = 32h$. For the case of the 8:1 contraction, after
onset of diverging flow, the long upstream tail of low, positive
strain rate did not extend as far upstream, reaching only to $\zeta$
= -5.5 instead of $\zeta \sim  -10$; a limited number of data points
prevents precise delineation of this change in extent.

The transition from a moderately peaked to a sharply peaked
strain-rate profile occurs smoothly over an intermediate range of
$We_{\mathrm{Up}}$. The maximum dimensionless strain rate is shown
as a function of $We_{\mathrm{Up}}$ in Figure 11(a) for the three
contraction ratios investigated. A square-root function of the form
\begin{equation}
{\dot{\epsilon}}_{\mathrm{max}} (We_{\mathrm{Up}})({h \over
{{\langle v_{z} \rangle}_{\mathrm{Dn}}}}) -
{\dot{\epsilon}}^{0}_{\mathrm{max}} ({h \over {{\langle v_{z}
\rangle}_{\mathrm{Dn}}}}) = C
(We_{\mathrm{Up}}-We_{\mathrm{Up,crit}})^{1/2}
\end{equation}
was fit to each data set associated with a given contraction ratio.
In equation (6), $We_{\mathrm{Up}}$ is the independent variable, and
${\dot{\epsilon}}^{0}_{\mathrm{max}}({h \over {{\langle v_{z}
\rangle}_{\mathrm{Dn}}}})$ is the maximum dimensionless centerline
strain rate in the limit of zero Weissenberg number, as inferred
from the data which is also used to fit the constants $C$ and
$We_{\mathrm{Up,crit}}$. Our experimental observations are consistent with a
supercritical bifurcation; however, there are not sufficient data
points for the bifurcation to be definitively classified. Thus,
these square-root fits are intended to act primarily as a guide to
the eye.

Two regions are apparent for the data associated with a given
contraction ratio. At low $We_{\mathrm{Up}}$, the maximum
dimensionless strain rate is nearly independent of
$We_{\mathrm{Up}}$; this independence is evident for
$We_{\mathrm{Up}} < 0.1$ for 8:1 contraction flow and for
$We_{\mathrm{Up}} < 0.02$ for 32:1 contraction flow. At larger
$We_{\mathrm{Up}}$ the maximum dimensionless strain rate increased
with $We_{\mathrm{Up}}$. Upstream diverging flow and downstream
velocity overshoot are noted in the velocity profile for elevated
volumetric flow rates; these phenomena result in the increase in
maximum dimensionless strain rate. Since the Weissenberg number is
defined in terms of the upstream flow parameters, the critical value
for transitions can be substantially less than unity. This reflects
the dependence of the flow rearrangement on contraction ratio as
well as on $We_{\mathrm{Up}}$.

The maximum dimensionless strain rate as a function of the
Weissenberg number defined in terms of the mean downstream shear
rate; i.e. $We_{\mathrm{Dn}} = We({\langle v_{z}
\rangle}_{\mathrm{Dn}}/h)$ is shown in Figure 11(b). The critical
parameters for the onset of diverging flow differ by only a factor
of 3 between $H/h = 2$ and $H/h = 32$, as opposed to the
corresponding parameters from Figure 11(a), which differ by a factor
of 20. However, distinct superposition of the curves is not
observed. This is attributed to the role which the contraction ratio
plays in governing the flow rearrangement. {\it It is clear that neither
the upstream nor downstream based Weissenberg number alone fully
describes the conditions for transition to diverging flow}.

To decide whether the upstream or downstream definition is the more
appropriate one to use, it is helpful to consider the critical
Weissenberg number for flow rearrangement expected in the limits of
$H/h \rightarrow 1$ (channel flow) and $H/h \rightarrow \infty$
(sink flow). For the sink-flow limit, it is expected that
$We_{\mathrm{Dn,crit}}$  is a non-zero, finite value. Specifically
as $H/h \rightarrow \infty$ , the upstream walls have less and less
effect on what occurs in the vicinity of the downstream slit. In
contrast, $We_{\mathrm{Up,crit}}$ will continuously decrease and
approach zero. In the channel-flow limit, it necessarily follows
that $\lim_{H/h \rightarrow 1}$
$We_{\mathrm{Up,crit}}$=$We_{\mathrm{Dn,crit}}$. In a channel,
rearrangement to diverging flow is never observed; two scenarios may
be envisioned as occurring as the lower limit is approached. 1) The
critical Weissenberg number may continuously approach infinity as
the contraction ratio approaches one. In this case
$We_{\mathrm{Up,crit}}$  will monotonically increase with $H/h$,
whereas $We_{\mathrm{Dn,crit}}$ will first decrease  and then
increase. 2) The critical Weissenberg number will approach a finite
limit as $H/h \rightarrow 1$ . $We_{\mathrm{Up,crit}}$ will
continuously increase, and $We_{\mathrm{Dn,crit}}$ continuously
decrease to the limit as channel flow is approached. By
consideration of the critical values shown in figures 11(a) and
11(b), this limiting value (should Scenario 2 apply) can be bounded
as $\lim_{H/h \rightarrow 1}$ $0.3 <
We_{\mathrm{Dn,crit}}$=$We_{\mathrm{Up,crit}} < 1.1$. Note, however,
that when the limit $H/h = 1$ is reached, the solution will vanish
since diverging flow does not occur in channel flow. Hence, as the
contraction ratio increase from the channel- to the sink-flow limit,
$We_{\mathrm{Up,crit}}$ will monotonically decrease, regardless of
whether Scenario 1 or Scenario 2 applies. This decrease reflects the
importance of both nonlinear elastic effects (as parameterized by
$We_{\mathrm{Up,crit}}$) and the imposed boundary conditions (as
parameterized by $H/h$) in determining rearrangement to diverging
flow; i.e. diverging flow is not noted in a channel of $H/h = 1$. In
contrast, $We_{\mathrm{Dn,crit}}$ increases with contraction ratio,
at least over a certain range of $H/h$. Furthermore,
$We_{\mathrm{Dn,crit}}$ will not necessarily exhibit a monotonic
dependence on $H/h$. These considerations motivate and support the use of the
Weissenberg number defined in terms of upstream flow conditions
throughout this paper.

\section{Interpretation of results using transition maps}

In the previous sections, qualitative flow visualization studies and
quantitative LDV measurements used to characterize the evolution of
the flow field with increasing $We_{\mathrm{Up,crit}}$ for a set of
planar contraction geometries were described. These results are
used now to develop a unified picture of velocity field transitions
in viscoelastic planar contraction flow. 
Characteristic length and time scales of
flow structures after the transitions are identified and related to
dimensions of the geometry. The viscoelastic scaling described in
the introduction is used to understand the onset of instability in
planar contraction flow in terms of the interaction of elastic
stresses oriented in the streamwise direction with streamline
curvature. Finally, a correlation between the onset of diverging
flow and global, spatial flow transition is noted.

\begin{figure}
\includegraphics[width=15cm]{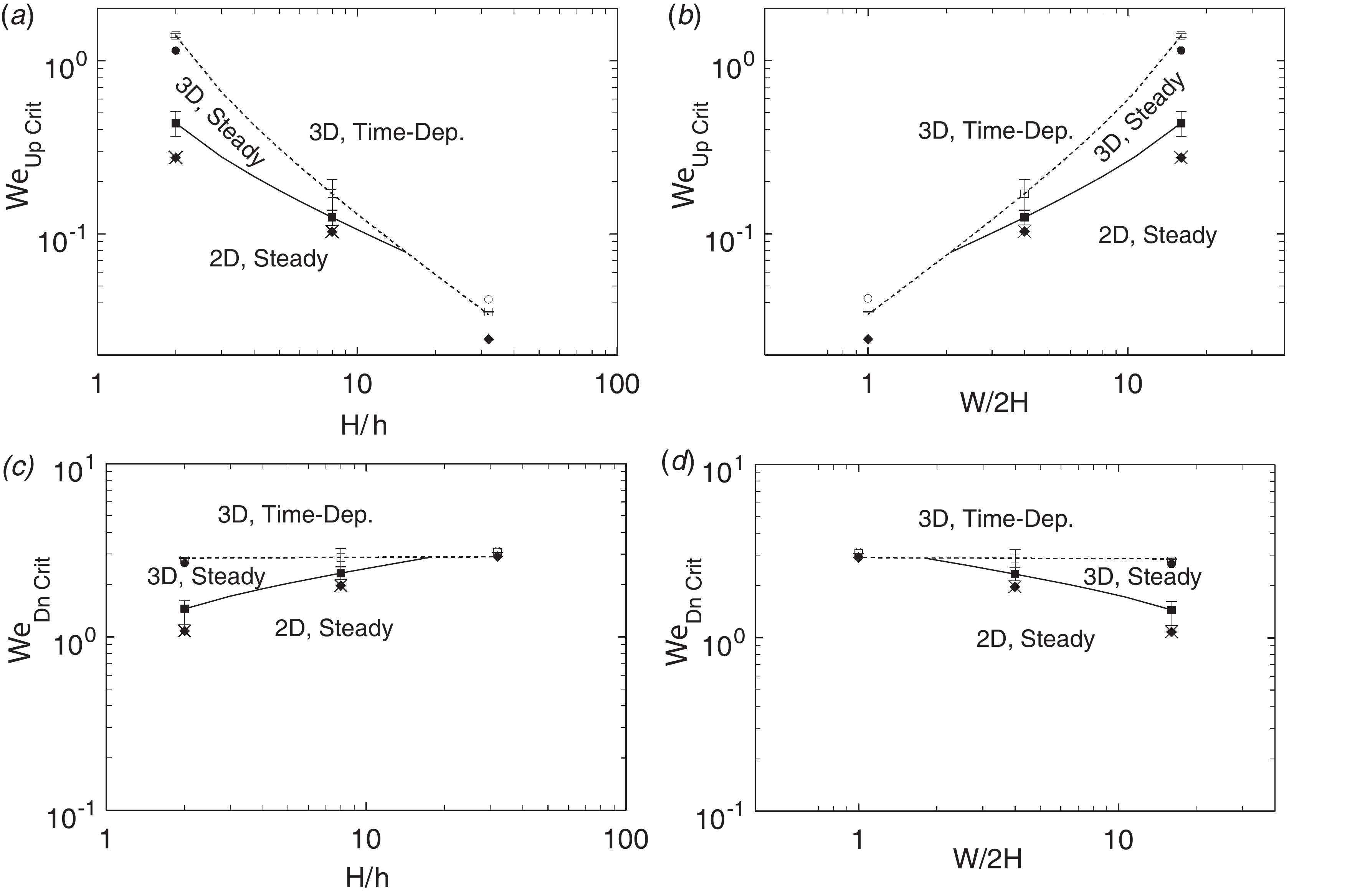}
\caption{Maps of transitions observed in viscoelastic planar
contraction flow: (a) critical $We_{\mathrm{Up}}$ vs. contraction
ratio, $H/h$; (b) critical $We_{\mathrm{Up}}$ vs. upstream aspect
ratio, $W/2H$. Experimentally observed transitions are denoted:
($\blacklozenge$), onset of increase in maximum dimensionless
centerline strain rate with $We_{\mathrm{Up}}$; ($\times$) diverging
flow; ($\blacksquare$) , transition from tow-dimensional, steady
base flow to three-dimensional, steady flow; ($\square$), transition
from steady to time-dependent flow; ($\bullet$), secondary spatial
harmonics; ($\bigcirc$) secondary spatio-temporal harmonics. Fits
are denoted: (solid line) , transition from two-dimensional, steady
base flow to three-dimensional, steady flow; (dashed line),
transition from steady to time-dependent flow; (c) critical
$We_{\mathrm{Dn}}$ vs. contraction
ratio, $H/h$; (d) critical $We_{\mathrm{Dn}}$ vs. upstream aspect
ratio, $W/2H$. Experimentally observed transitions are denoted:
($\blacklozenge$), onset of increase in maximum dimensionless
centerline strain rate with $We_{\mathrm{Dn}}$; ($\times$) diverging
flow; ($\blacksquare$), transition from two-dimensional, steady base
flow to three-dimensional, steady flow; ($\square$), transition from
steady to time-dependent flow; ($\bullet$), secondary spatial
harmonics; ($\bigcirc$) secondary spatio-temporal harmonics. Fits
are denoted: (solid line) , transition from two-dimensional, steady
base flow to three-dimensional, steady flow; (dashed line),
transition from steady to time-dependent flow.}
\end{figure}

\subsection{Flow transition maps}

The series of ordered flow transitions denoted by the critical value
of $We_{\mathrm{Up}}$ are shown in Figures 12(a) and 12(b) as a
function of the contraction ratio, $H/h$ and the upstream aspect
ratio. It is clear from Figure 12(a) that the 2:1 and 8:1
contractions exhibit a common transition sequence: two-dimensional
rearrangement to diverging flow, transition from two-dimensional to
three-dimensional and steady flow, and onset of time-dependent flow.

\subsubsection{Transitions in the 2:1 and 8:1 Contractions}

For $H/h = 2$ and 8 the flow is two-dimensional and steady at low
values of $We_{\mathrm{Up}}$. As $We_{\mathrm{Up}}$ is increased
above a lower critical value, the centerline velocity profile
changes. The flow begins to diverge upstream of the contraction
plane, and velocity overshoot is observed downstream of the
contraction plane; both phenomena are associated with an increase in
the maximum dimensionless centerline strain rate. As the flow rate
is increased above the lower critical value, a subsequent
supercritical bifurcation to three-dimensional and steady flow
throughout the upstream region was observed. For the 2:1
contraction, wavenumber doubling behavior was detected at yet higher
values of $We_{\mathrm{Up}}$.

The next transition observed with increasing volumetric flow rate
was a bifurcation to time-dependent flow, corresponding to a
supercritical Hopf bifurcation. As discussed earlier the amplitude
of the temporal oscillation of $v_{z}$ grew continuously and
monotonically with $We_{\mathrm{Up}}$, consistent with the scaling
of a supercritical Hopf bifurcation. The flow pattern close to
criticality has temporal oscillations (traveling waves) superposed
on the spatial oscillation (stationary waves). To understand this
flow structure, consider the appearance of the time series if there
were no superposition and the waves associated with the spatial
oscillation suddenly started to move as a unit from the bounding
wall toward the center of the flow. A sudden jump, not a gradual
increase, in the amplitude of temporal oscillation would be observed
at a given point in space. Videotaped streakline images in the
$xz$-plane of the 2:1 contraction were studied; there were
indications that the spatio-temporal structure after onset of the
temporal instability was more complex than that of a single set of
waves traveling at a uniform velocity in the $x$-direction. However,
the precise spatio-temporal structure could not be determined from
these streakline images.

The flow visualization observations of the temporal instability for
the 8:1 contraction indicated that the vortex bundles traveled from
the walls of the geometry toward the center of the flow; no
underlying stationary wave pattern was observed. However, streakline
visualization may not detect low amplitude oscillations near onset.
Only images of the central region of the flow  could be acquired.
Since the amplitude of the temporal oscillation was greatest near
the bounding walls, the flow visualization technique may not have
been capable of resolving superposed temporal and spatial waves near
onset of the temporal instability. The observation via light sheet
visualization in the 8:1 contraction of moving streamline bundles
without an underlying stationary pattern, is probably a result of
the $We_{\mathrm{Up}}$ being substantially greater than the critical
value for onset of temporal instability, so that large amplitude
traveling waves dominated over low amplitude stationary waves.

\subsubsection{Flow transitions in the 32:1 contraction}

For flow through the 32:1 contraction, we found that at low values
of $We_{\mathrm{Up}}$, the flow was steady and two-dimensional. As
$We_{\mathrm{Up}}$ was increased above a critical value, an increase
in the maximum dimensionless centerline strain-rate was observed.
However, diverging flow was not seen; the strain rate was positive
at all points on the centerline upstream of the contraction plane.

At still higher values of $We_{\mathrm{Up}}$, a direct transition
from globally steady, two-dimensional flow to a three-dimensional,
time-dependent flow was found. Two scenarios could explain this
observation. One, the width of the geometry may have been too narrow
for the steady, three-dimensional flow, characterized by spatial
oscillation of $v_{z}$ in the $x$-direction, to occur. Alternately,
it also is possible that onset of the temporal flow transition is
the result of an interaction between the local flow transition
noticed at the bounding wall and the global flow field. For high
upstream aspect ratios, the local vortex at the wall would have to
grow substantially, requiring higher values of $We_{\mathrm{Up}}$,
before interacting with the entire flow field. In contrast, for low
upstream aspect ratio, only a small amount of growth of this
bounding wall vortex would be required, resulting in low values of
$We_{\mathrm{Up}}$ for onset of the temporal instability. Hence, for
the 32:1 contraction the extreme case of unit aspect ratio induces
onset of the temporal oscillations before onset of steady, spatial
oscillations. The latter view is supported by LDV measurements which
indicated that the amplitude of the temporal oscillation was
greatest near the bounding wall at $\chi = -32$ and weakest near the
center of the flow $\chi =0$. At high flow rates, the onset of
harmonics in the frequency spectrum was noted. These harmonics
seemed to occur by the spatial wavenumber doubling of the wave
structure and not doubling of the rate of wave propagation in the
$x$-direction.

\subsubsection{Summary of the transition maps}

The maps illustrate that there are commonalities in the sequence of
transitions which occur for increasing $We_{\mathrm{Up}}$ for
different contraction ratios; however, the critical values of
$We_{\mathrm{Up}}$ for these transitions decrease with increasing
contraction ratio. The ratio of critical onset values,
$We_{\mathrm{Up,T}}/We_{\mathrm{Up,S}}$ decreases between the
contraction ratios 2:1 and 8:1 and apparently decreases to a value
less than unity for the 32:1 contraction.

The critical value for onset of the spatial instability,
$We_{\mathrm{Up,S}}$, appears to be directly related to the ratio of
upstream to downstream half-height, $H/h$. We  address this issue later. 
The onset of the temporal flow
transition also appears to be closely related to the upstream aspect
ratio, $W/2H$.

\subsubsection{Transition maps in terms of downstream Weissenberg numbers}

The transition map plotted in Figures 12(c) and 12(d) shows the
critical Weissenberg number defined in terms of downstream flow
parameters, $We_{\mathrm{Dn,crit}}$, as a function of contraction
ratio or aspect ratio. This quantity defined in terms of the mean
downstream shear rate for the spatial transition,
$We_{\mathrm{Dn,S}}$, increases by a factor of two as $H/h$
increases from 2 to 8. This behavior may be contrasted with the
four-fold decrease of $We_{\mathrm{Up,S}}$ for the same increase in
$H/h$. Thus, when the range is defined in terms of
$We_{\mathrm{Dn,S}}$, the curve associated with the spatial
transition appears flatter than when the upstream Weissenberg number
is used. Note, however, that whereas $We_{\mathrm{Up,S}}$ decreases
with contraction ratio, $We_{\mathrm{Dn,S}}$ increases with
contraction ratio. In subsequent sections, the relation of the
critical Weissenberg number for the spatial transition to the
streamline curvature around the outer corner which is induced by the
greater-than-unity contraction ratio, $H/h > 1$, is discussed.
Consideration of the streamline-curvature interaction leads to the
prediction that for increased curvature the critical Weissenberg
number will decrease. The observation that the Weissenberg number
decreases when defined in terms of upstream parameters  but not when
defined in terms of downstream parameters supports the use of the
upstream Weissenberg number.

The critical value for transition from steady to time-dependent
flow, $We_{\mathrm{Dn,T}}$, appears insensitive to contraction
ratio. In contrast, the value of $We_{\mathrm{Up,T}}$ for the
temporal transition exhibits a strong dependence on $H/h$.
Superficially, this observation appears to indicate that the
transition to time-dependent behavior is controlled by the parameter
$We_{\mathrm{Dn}}$. However, in fact, the temporal transition
appears to be related to flow conditions induced by the presence of
a wall bounding the lateral sides of the upstream channel. In this
scenario, an appropriate Weissenberg number could be defined in terms
of the characteristic shear rate in the region near the centerplane,
immediately adjacent to the wall bounding the x-dimension, and close
to the contraction plane. Note that the width of the channel, $W$,
and the half-height of the downstream channel, $h$, were fixed for
all the experiments. Hence the characteristic shear rate alluded to
is approximately proportional to the mean downstream shear rate. The
flatness of the curve describing the temporal transition does not
necessarily imply that downstream flow conditions control the
spatial transition. Rather, this flatness may be related to a
critical value of the characteristic shear rate and the particular
geometry used for these experiments, in which the ratio of width to
downstream half-height was constant at $W/2h = 32$.

\subsection{Characteristic length and time scales of flow
structures}

Characteristic length scales associated with the flow rearrangement
to diverging flow and the spatial transition to three-dimensional
flow were estimated from the measurements. These length scales were
in turn related to geometric parameters of the planar contraction. A
characteristic time scale was extracted for the temporal
oscillation; however, this time scale could not be related to
specific parameters of the flow or fluid rheology (e.g. the
parameters characterizing the relaxation spectrum, $\lambda_{k}$).
In addition, a length scale associated with the diverging flow was
obtained by determining the distance of the point of minimum
velocity on the centerline from the contraction plane at $\zeta =
0$.

The wavenumber of the spatial oscillation after flow transition,
made dimensionless with the upstream half-height, is plotted in
Figure 13(a) as a function of $We_{\mathrm{Up}}$ for the 2:1 and 8:1
contraction ratios. For the 8:1 contraction, the dimensionless
wavenumber increased with $We_{\mathrm{Up}}$; for the 2:1
contraction, no significant dependence of wavenumber on
$We_{\mathrm{Up}}$ was noted. Above a critical value of
$We_{\mathrm{Up}}$, but before onset of time-dependent flow,
doubling of the spatial wavenumber in the 2:1 contraction was
observed. Period doubling in the 8:1 contraction was not observed.
The dimensionless wavenumber of the primary peak is of the same
order for both the 2:1 and 8:1 contraction ratios, $H/\lambda_{x}
\approx 0.7$. A three-dimensional and steady flow state was not seen
in flow through a 32:1 contraction.

\begin{figure}
\includegraphics[width=12cm]{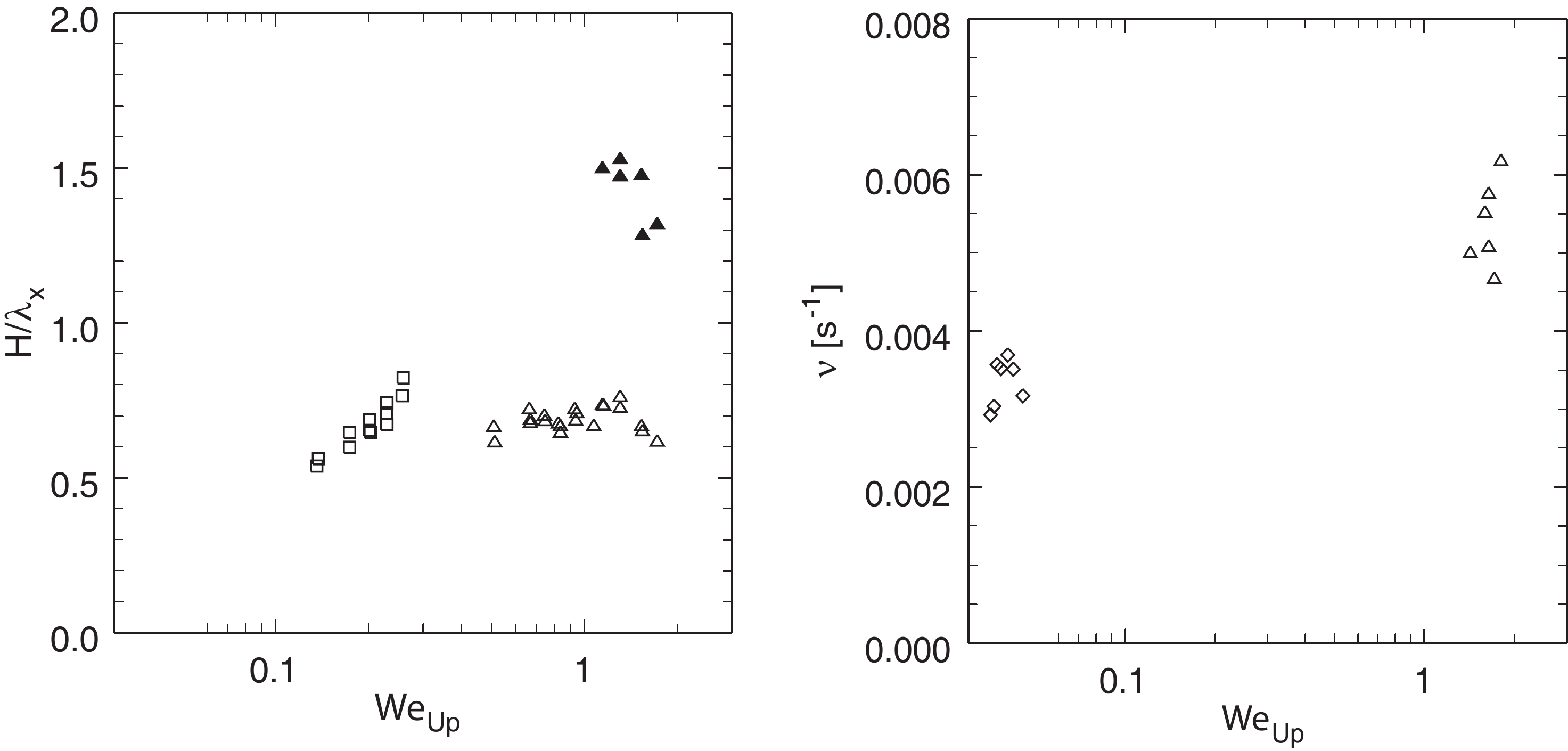}
\caption{ (a) Dimensionless wavenumber, $H/\lambda_{x}$, vs.
$We_{\mathrm{Up}}$ after onset of three-dimensional flow:
($\square$) primary peak for the 8:1 contraction, streakline
visualization data used; primary ($\bigtriangleup$) and secondary
($\blacktriangle$) peaks for the 2:1 contraction, LDV data used,
scans performed over range $-26 \leq \chi \leq -1.5$,  $\upsilon$ =
-1.75, $\zeta$ = -1.80.
(b) Frequency ($\nu$ ) vs. $We_{\mathrm{Up}}$ after onset of
time-dependent flow: ($\triangle$), 2:1 contraction, $v_{z}$ vs. $t$
series acquired at $\chi$ = -20.0, $\upsilon$ = -1.75, $\zeta$ =
-1.80; ($\lozenge$) , 32:1 contraction, $v_{y}$ vs. $t$ series
acquired at $\chi$ = -21.0, $\upsilon$ = -1.51, $\zeta$ = -1.51. }
\end{figure}

The characteristic frequency of the temporal oscillation is shown in
Figure 13(b) as a function of $We_{\mathrm{Up}}$ for the 2:1 and 32:1
contraction ratios. The time-series data used to determine this
frequency was acquired with the LDV measuring volume placed
approximately half-way between the center of the flow and the
bounding wall of the channel: at  $\chi =  -20$ for the 2:1
contraction, and at  $\chi =  -21$ for the 32:1 contraction. The
results indicate only a weak dependence of frequency on contraction
ratio: the frequency of oscillation in the 2:1 contraction is higher
than the frequency in the 32:1 contraction. Note that the period
associated with the oscillation was $T_{p} = 180$ s for the 2:1
contraction and $T_{p}  = 340$ s for the 32:1 contraction. In both
cases, the period is much greater than the estimated zero-shear-rate
relaxation time of the test fluid, $\lambda_{10} =
\psi_{10}/2\eta_{o} \sim 1$ s. No trend of the frequency was
observed with variation of $We_{\mathrm{Up}}$, for a given
contraction ratio. LDV measurements were not taken for the 8:1
contraction; data was obtained via analysis of videotaped streakline
images.

\subsection{Critical Weissenberg numbers
and mechanism driving transitions}

Consideration of structural features of the flow transitions and the
scaling of onset $We_{\mathrm{Up}}$ with contraction ratio has
elucidated the mechanism of viscoelastic instability in the planar
contraction. In particular, these features were consistent with the
interaction of streamwise elastic stress with streamline curvature
around the outer corners of the planar contraction acting to induce
flow transitions.

Shear flow rheological characterization of the Boger test fluid
indicated that the fluid had a high zero-shear-rate relaxation time
of $\lambda_{10} > 1$ s. Inertia was negligible, $Re_{\mathrm{Dn}} =
7 \times 10^{-4}$, for all the flows investigated, whereas elastic
memory or nonlinear effects could influence the flows, since
$We_{\mathrm{Up}}$ as great as 1.8 were attained. Therefore,
although some structural features of the flow after onset of
instability resemble those of G\"{o}rtler vortices, the transitions
were induced by the elastic, not inertial, character of the flows.
The characteristic length scales of flow structures indicate that
the transitions were driven by interaction of elastic stresses in
the streamwise direction with streamline curvature in the upstream
region of the flow. Specifically, the extent of the oscillations
associated with the flow transitions upstream of the contraction
plane, as indicated by the reattachment length of the outer corner
vortex adjacent to the fast flow region, was of the order of the
upstream half-height, $H$. The wavelength of the spatial oscillation
was quantitatively characterized via LDV scans in the neutral
$x$-direction and found to scale with upstream half-height. The
characteristic dimensions of the oscillation in the $x$- and in the
$z$-directions indicate the relation of the flow transition to
streamline curvature in the flow around the outer corner, rather
than around the reentrant corner.

The spatial structure of the flow after onset of the instability
bears resemblance to viscometric flows for which it has been
concluded that interaction of streamwise normal stresses with curved
streamlines drives the instability. To unify previous experimental
observations of flow transitions, McKinley and co-workers [10]
introduced and used the idea of transition maps. Employing physical arguments
and scalings based on the linearized form of the governing
equations, a dimensionless parameter was formulated that related a
characteristic shear rate, streamwise stress, fluid residence time,
and streamline curvature in a flow to a viscoelastic G\"{o}rtler
number, $M_{VG}$, given by
\begin{equation}
M_{VG} \equiv ({{V \lambda_{p1} \over R_{c}}} {{\tau_{p11} \over
{\eta \dot{\gamma}}}})^{1/2}.
\end{equation}
This dimensionless quantity was found to set the conditions for the
onset of instability for several viscometric and complex flows. The
first term within the parentheses represents a ratio of a
characteristic length over which perturbations to the viscoelastic
stress relax to a characteristic radius of curvature in the flow,
$R_c$. The relaxation length in the numerator is the product of a
characteristic velocity $V$ and the characteristic relaxation time
of the polymer, $\lambda_{p1} \equiv \psi_{1}/2(\eta-\eta_{s})$
where $\psi_{1}$ is the first normal stress coefficient, $\eta$ is
the solution viscosity, and $\eta_{s}$ denotes the solvent
viscosity. The second term within the parentheses represents the
relative magnitude of the coupling of perturbative elastic stresses
to the stresses in the base flow. The polymeric contribution to the
streamwise stress is given by  $\tau_{p11}$, and $\dot{\gamma}$  is
a characteristic shear rate. For complex flows, parameters that
relate the characteristic streamline curvature to geometric
parameters must be computed or measured.

To understand the critical Weissenberg numbers observed in our
experiments in more detail, we start with the relationship embodied in
equation (7). The upstream half-height,
$H$, is the characteristic length scale for the extent of the
spatial oscillation upstream of the contraction plane and the
wavelength of the oscillation; therefore, the flow transition seems
to be related to streamline curvature around the outer corner of the
contraction. Hence, it is natural to use the upstream flow
conditions in obtaining characteristic values.

The characteristic curvature, $R_{c}^{-1}$, must approach zero as
the contraction ratio approaches unity; this limit is equivalent to
fully developed Poiseuille flow in a channel for which
elastically-driven instabilities are not observed. As $H/h$
increases, the dimensionless curvature, $H/R_{c}$, of the
streamlines around the outer corner of the contraction is expected
to increase. A simple expression that captures this behavior is
\begin{equation}
{H \over R_{c}} = {1 \over A}({H \over h} - 1)^{B}
\end{equation}
where the constants $A$ and $B$ are both positive. The velocity,
$U$, and shear rate, $\dot{\gamma}$, are related to upstream
flow conditions as
\[
\dot{\gamma} = {\langle \dot{\gamma} \rangle}_{\mathrm{Up}} = U/H =
Q(2H^{2}W)^{-1},
\]
where $W$ is the extent in the $x$-direction of the geometry and $Q$, the
volumetric flow rate. This shear rate is used in the evaluation of
the polymeric relaxation time,  $\lambda_{p1}$, and the streamwise
polymeric stress, $\tau_{p11}$. The expression for the critical
condition for onset of flow instability in the planar contraction is
re-written in terms of a critical upstream shear-rate-dependent
Weissenberg number as
\begin{equation}
We_{\mathrm{Up,crit}} = M_{VG}({A(1-\eta_{s}/\eta) \over
2})^{1/2}({H \over h}-1)^{-B/2} = S ({H \over h}-1)^{-B/2}
\end{equation}
where the unknowns $A$ and $M_{VG}$ have been combined into a single
term, $S = M_{VG}({A(1-\eta_{s}/\eta)/2})^{1/2}$. Experimental data
for the critical upstream Weissenberg number, $We_{\mathrm{Up,S}}$,
associated with the spatial flow transition in the 2:1 and 8:1
contractions are used to fit the unknown parameters. Specifically, a
good match with the data for spatial transition was obtained with
$S_{S} = 0.44$ and $B_{S}/2 = 0.64$.

The conditions governing transition to time-dependent flow are more
complex than those controlling onset of three-dimensional, steady
flow. As mentioned earlier, a local flow transition near the wall
bounding the $x$-dimension seems to play a role in the onset of the
temporal transition; the vortex structure associated with this local
transition scales with the downstream half-height, $h$, rather than
the upstream half-height, $H$. The upstream aspect ratio, $W/2H$, in
addition to the contraction ratio, $H/h$, together determine the
$We_{\mathrm{Up,T}}$ for transition to time-dependent flow.
Nevertheless, we obtain a reasonably good fit to the data for the
critical $We_{\mathrm{Up}}$ for transition to time-dependent flow in
the geometries with contraction ratio $H/h$ = 2, 8, and 32. The
fitted parameters had values of $S_{T} = 1.4$ and $B_{T}/2 = 1.1$;
the root mean square of the fractional error of the fit was 3\%.
These fits have been plotted along with experimental data in the
figures illustrating the flow transition.

The conditions governing transition to time-dependent flow may be
more complex than those controlling onset of three-dimensional,
steady flow. As discussed earlier, a local flow transition near the
wall bounding the $x$-dimension seems to play a role in the onset of
the temporal transition; the vortex structure associated with this
local transition scales with the downstream half-height, $h$, rather
than the upstream half-height, H. The upstream aspect ratio, $W/2H$,
in addition to the contraction ratio, $H/h$, may determine the
$We_{\mathrm{Up,T}}$ for transition to time-dependent flow.
Nevertheless, equations (8) and (9) provide a good fit to the data for the
critical $We_{\mathrm{Up}}$ for transition to time-dependent flow in
the geometries with contraction ratio $H/h$ = 2, 8, and 32. The
fitted parameters had values of $S_{\mathrm{T}} = 1.4$ and
$B_{\mathrm{T}}/2$ = 1.1; the root mean square of the fractional
error of the fit was 3\%.

When results for the critical Weissenberg numbers for the transition
data are extrapolated to higher contraction ratios, the neutral
stability curves for the predicted spatial (two-dimensional to
three-dimensional flow) and temporal (steady to time-dependent flow)
transitions intersect at a contraction ratio of approximately
$(H/h)_{S} = 15$. In the experiments, the contraction ratio was
adjusted by varying the upstream channel half-height for a fixed
downstream half-height. Hence, an increase in the contraction ratio
results in a decrease in the upstream aspect ratio. Our results
suggest that the point of intersection of the neutral stability
curves corresponds to an upstream aspect ratio of $(W/2H)_{S} =
2.1$. The dimensionless wavenumber for the 2:1 and 8:1 contractions
was determined to be approximately $H/\lambda_{x} = 0.7$. This
provides an estimate of the number of wave cycles that would fit
into the half-width of a geometry with an upstream aspect ratio of
two: $W(2\lambda_{x})^{-1} = 1.5 \sim 1$. It is interesting that
approximately one wave cycle of the spatial oscillation would fit
into the half-width of the geometry with a contraction ratio such
that $We_{\mathrm{Up}}$ is identical for onset of the spatial and
temporal instabilities. This strongly supports the idea that
transition to time-dependent flow is related to a three-dimensional
imperfection introduced by the wall bounding the lateral dimension
of the geometry. In particular, the base flow may make a direct
transition to time-dependent flow in the 32:1 contraction since a
full cycle of the spatial oscillation cannot fit into the half-width
of the upstream channel.

As discussed earlier, for the 2:1 and 8:1 contractions, onset of
three-dimensional, steady flow occurs for a $We_{\mathrm{Up}}$
immediately greater than the $We_{\mathrm{Up}}$ for which an
increase in the maximum dimensionless centerline strain rate with
$We_{\mathrm{Up}}$ is noted. A global transition to
three-dimensional flow, distinct from the transition to
time-dependent flow, was not noted for the case of the 32:1
contraction. However, the critical $We_{\mathrm{Up}}$ for direct
transition from two-dimensional, steady to three-dimensional,
time-dependent flow in the 32:1 contraction is greater than the
critical $We_{\mathrm{Up}}$ for onset of increasing centerline
maximum dimensionless strain rate with $We_{\mathrm{Up}}$. Hence,
the relation between flow transition and the dependence of peak
strain rate on $We_{\mathrm{Up}}$ for the case of the 32:1
contraction is consistent with the relation observed for the 2:1 and
8:1 contractions.

It is possible that the two-dimensional flow rearrangement
associated with the diverging flow promotes onset of the
three-dimensional and steady instability. In particular,
visualization of flow in the $yz$-plane indicates that after onset
of diverging flow, the streamlines near the outer corner become more
tightly curved. A smaller characteristic radius of curvature of the
streamlines in conjunction with an increase in the local strain
rate, associated with shift of the streamlines toward the
contraction plane, may allow the critical viscoelastic G\"{o}rtler
number, for onset of the instability, to be exceeded for a lower
volumetric flow rate than if this two-dimensional streamline
rearrangement did not first occur. This could explain why the
critical Weissenberg number for transition to three-dimensional flow
is immediately greater than the critical $We_{\mathrm{Up}}$ for
onset of diverging flow.

\section{Future perspectives and concluding remarks}

Our experiments support the concept that the interaction of
stream-wise stress with streamline curvature induces transition to
three-dimensional flow in complex flows, such as in the planar
contraction.  Two interesting questions are yet to be understood satisfactorily.
First, more detailed information on the stress-curvature mechanism
which drives the transition to three-dimensional flow is required.
Secondly, the influence of three-dimensionality in the base flow,
introduced by the wall bounding the neutral $x$-dimension, on the
flow transition sequence is yet to be understood.
These
walls are seen to act as an imperfection to the nominally two-dimensional
base flow and alter the essential nature of the flow transition
sequence.

There is evidence that a bifurcation of codimension-2 [18,19] exists
at the point where the neutral stability curves for the spatial and
the temporal transitions intersect on the transition map; this point
occurs at approximately $H/h = 15$ (or $W/2H = 2.1$) and
$We_{\mathrm{Up}} = 0.08$. To achieve such a bifurcation two control
parameters, in this case $H/h$ (or $W/2H$) and $We_{\mathrm{Up}}$,
must be tuned to specific values.
The transition from the two-dimensional, steady base
flow in the planar contraction appears to be driven by the
inter-action of stream-wise stress with streamline curvature. This
implies that such a transition would be observed in an ideal,
two-dimensional planar contraction flow. An imperfection in the form
of finite, but large, upstream aspect ratio might modify this
transition but would not be expected to change its basic structure.

The upstream
aspect ratio, $W/2H$, is observed to play an essential role in defining the
critical conditions for transition from steady to time-dependent
flow but not for the spatial transition. This suggests that the
transition map for viscoelastic flow through a planar contraction of
finite upstream aspect ratio is most appropriately represented in a
three-parameter space. The upstream Weissenberg number, the upstream
aspect ratio $(W/2H)$, and the contraction ratio $(H/h)$ are the
independent parameters that determine the flow state of the system
and are each associated with a coordinate axis of the space. Figure
14 shows a qualitative transition map in this three-parameter space
for flow through the planar contraction. The neutral stability
boundary for the spatial transition from two-dimensional, steady
base flow to three-dimensional, steady flow is represented as a
two-dimensional surface within this space. In Figure 14, it is
assumed that this {\em spatial transition surface} is nearly
independent of aspect ratio. The neutral stability boundary
identified with the steady to time-dependent temporal transition is
represented as a second surface. This {\em temporal transition
surface} is also assumed to be independent of contraction ratio to
leading order. The intersection of the two neutral-stability
surfaces defines a one-dimensional curve on which a codimension-2
bifurcation occurs.

\begin{figure}
\includegraphics[width=8cm]{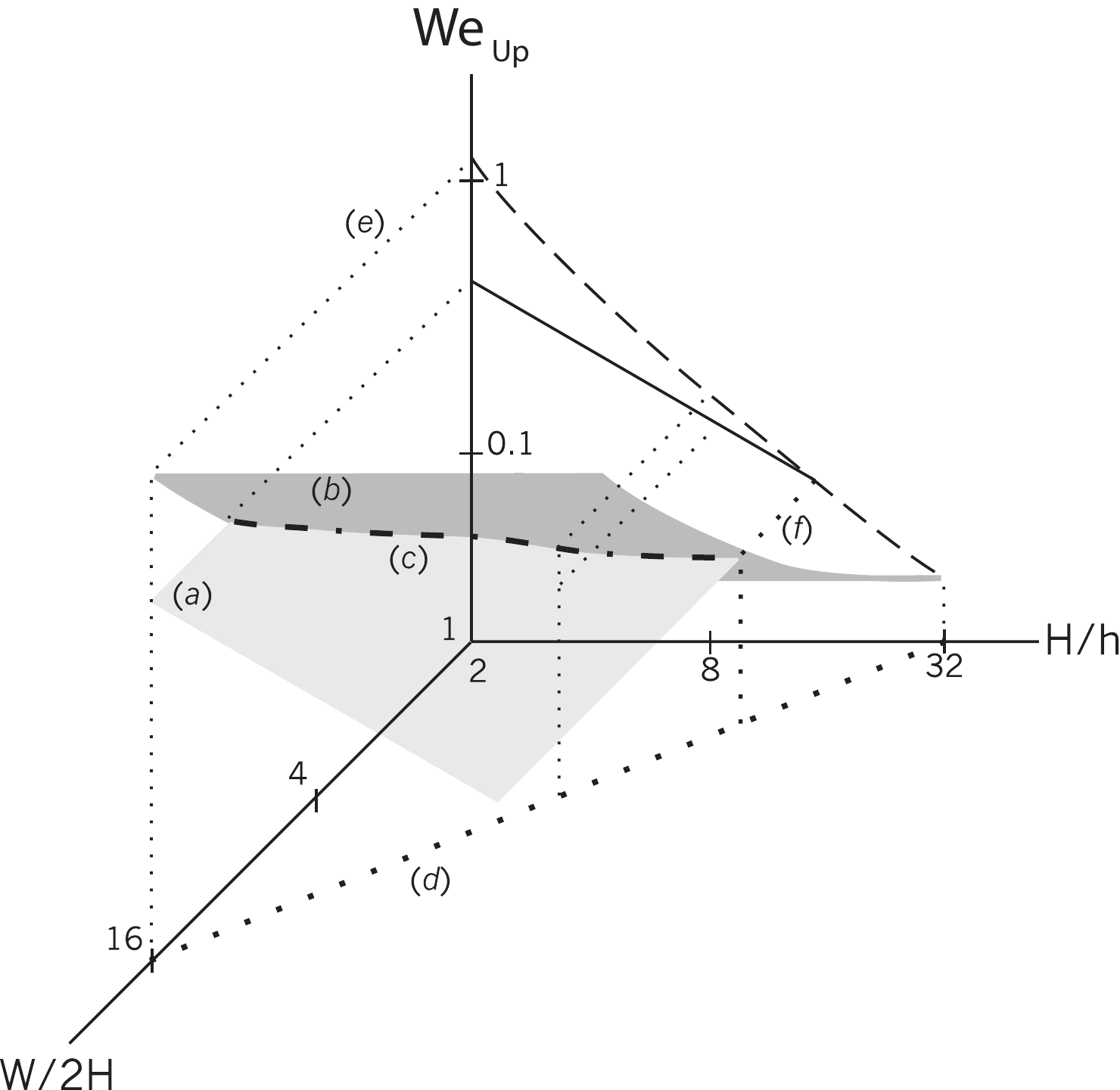}
\caption{Hypothetical representation of flow transition map for the
planar contraction in three-parameter space: ((a), light gray plane)
neutral-stability surface for spatial transition from two- to
three-dimensional steady flow; ((b), dark gray plane)
neutral-stability surface for temporal transition from steady to
time-dependent flow; ((c), $-.-$) curve for codimension-2
bifurcation at the intersection of the neutral stability surfaces;
((d), $---$) projection of the experimental subspace onto the
($W/2H,H/h$) plane; construction lines indicating the projection
onto the ($We_{\mathrm{Up}},H/h$) plane of the intersection of the
experimental subspace with neutral stability surfaces ((e), $....$)
and with the curve representing the codimension-2 bifurcation
((f),$----$ ).}
\end{figure}

In the experiments presented here, the upstream aspect ratio, $W/2H$,
and the contraction ratio, $H/h$, were simultaneously varied. The
relation between the two parameters viz., the upstream and
downstream aspect ratios are related to the contraction ratio simply
by $W/2H = (W/2h)/(H/h)$. The downstream aspect ratio was constant
throughout the experiments (with value $W/2h = 32$). These
relationships enable us to  define a two-dimensional {\em
experimental subspace}, which can be represented as a surface
(invariant along the $We_{\mathrm{Up}}$ axis) in the
three-dimensional transition map. The intersection of the
experimental subspace with the given neutral-stability boundary
 is the one-dimensional subset of the boundary
was probed in our experiments. Projection of the one-dimensional
neutral-stability subsets onto the $(We_{\mathrm{Up}},H/h)$ and
$(We_{\mathrm{Up}},W/2H)$ planes yields the two-dimensional
transition map.

The concept of a three-parameter transition map discussed above is
consistent with the experimental results but admittedly speculative.
Moreover, the form of the full three-parameter transition map
outside of the {\em experimental subspace}  can only be supposed.
Nevertheless, the concept of a three-parameter transition map
provides guidance for the design of future experiments.
Because of the
choice of the form of the two hypothetical neutral-stability
surfaces, their intersection, i.e. the curve associated with the
codimension-2 bifurcation, does not have $W/2H$ constant.
With this caveat, let us now consider the ramifications of such a
transition map. The experiments indicate that a codimension-2
bifurcation is likely to exist in the vicinity of the point in
parameter space with coordinates $We_{\mathrm{Up}} = 0.08$, $H/h =
15$, and $W/2H = 2.1$. The value of the upstream aspect ratio,
$W/2H$, is interesting in that approximately one cycle of the
spatial oscillation could fit into the half-width of the geometry,
$W/2$. A dynamical system in the vicinity of a codimension-2
bifurcation can exhibit behavior which is essentially different from
that observed near the individual neutral stability curves which
intersect at the codimension-2 point. It is well established that a
nonlinear dynamical system may behave quasi-periodically or even
chaotically when control parameters are nearly equal to values
associated with the codimension-2 bifurcation, even if such behavior
does not occur for either of the individual neutral stability
curves. In this context, it would be interesting to determine
whether the one-dimensional curve defining the location of the
codimension-2 bifurcation lies entirely within (or near to) the
plane in parameter space defined by $W/2H = 2$. In addition to these
sets of experiments, designed to map the transitions in parameter
space, detailed characterization of the flow dynamics should be
conducted for parameters in the vicinity of the codimension-2
bifurcation. Of particular interest is whether the dynamics are
essentially different from behavior observed on the neutral
stability surfaces where the bifurcation has codimension-1.

Substantial work is still required before a unified framework for
understanding the nature of transitions induced by the interaction
of stream-wise elastic stress and streamline curvature can be
constructed. A more detailed understanding of the mechanisms driving
flow transitions in complex geometries will most certainly require
numerical investigations. The increasing availability of
computational resources and the development of more efficient and
accurate algorithms [20-25] make such simulations feasible. In such
a scenario, we envisage the use of experimental data such as that
presented in this paper to serve both as a guide and as a benchmark
for numerical computations.

\section{Remarks in Conclusion}
Since the preparation of this manuscript in 2003/2004, 
significant experimental and computational work on  flow through 
contractions in microfluidic geometries and microchannel stagnation flows has been published
in the literature - see for instance Soulages et.al., JNNFM 163, 9-24 (2009),
Rodd et. al., JNNFM 143(2-3), 170-191 (2007), Rodd et. al., JNNFM, 129(1), 1-22 (2005).
Readers are referred to these publications and references therein for more recent investigations on the nature of flows through contraction geometries.

\section*{Acknowledgements}

Funding to support this research was provided by the National
Science Foundation. AG thanks Dr. Udupi Krishna for helpful discussions.

\section*{References}

[1]  Boger, D.V., Hur, D.U. and Binnington, R.J., Further
Observations of Elastic Effects in Tubular Entry Flows, {\it J
Non-Newtonian Fluid
  Mech.},
  {\bf 20} (1986) 31-49.

[2]  Binding, D. M. and Walters, K.,
 On the use of flow through a contraction in estimating extensional
 viscosity of mobile polymer solutions, {\it J. Non-Newtonian Fluid
 Mech.},
 {\bf 30} (1988) 233-250.

[3]  Evans, R.E. and Walters, K., Further Remarks on the Lip-Vortex
 Mechanism of Vortex Enhancement in Planar-Contraction Flows,
 {\it J. Non-Newtonian Fluid Mech.}, {\bf 32} (1989) 95-105.

[4]  Chiba, K., Sakatani, T. and Nakamura, K., Anomalous
 Flow Patterns in Viscoelastic Entry Flow through a Planar
 Contraction,
  {\it J. Non-Newtonian Fluid Mech.}, {\bf 36} (1990)193-203.

[5]  Larson, R.G., Shaqfeh, E.S.G. and Muller, S.J.,
 A Purely Elastic Instability in Taylor-Couette Flow, {\it J. Fluid
 Mech.},
 {\bf  218} (1990) 573-600.

[6]  Chiba, K., Tanaka, S. and Nakamura, K.,
 The Structure of Anomalous Entry Flow Patterns through a
 Planar Contraction, {\it J. Non-Newtonian Fluid Mech}., {\bf 42}
 (1992)
 315-322.

[7]  Joo, Y.L. and Shaqfeh, E.S.G., A Purely Elastic Instability
 in Dean and Taylor-Dean Flow, {\it Phys. Fluids} A, {\bf 4}(3) (1992)
 524-543.

[8]  Joo, Y.L. and Shaqfeh, E.S.G., Observations of Purely
 Elastic Instabilities in the Taylor-Dean Flow of a Boger Fluid,
 {\it J. Fluid Mech.}, {\bf 262} (1994) 27-73.

[9]  Baumert, B. M. and Muller, S. J.,
 Flow visualization of the elastic Taylor-Couette instability in
 Boger fluids,
 {\it  Rheol. Acta.}, {\bf  34} (1995) 147-159.

[10]  McKinley, G.H., Pakdel, P. and Ozetkin, A.,
 Geometric and rheological scaling of purely elastic flow
 instabilities,
 {\it J. Non-Newtonian Fluid Mech.}, {\bf 67} (1996) 19-48.

[11]  Nigen, S. and Walters, K., Visco-elastic contraction flows:
comparision of axisymmetric and planar configurations,  {\it J.
Non-Newtonian Fluid Mech.},  {\bf 102} (2002) 343-359.

[12]  Alves, M. A., Pinho, F. T., and Oliveira, P. J.,
Visualizations of viscoelastic flow in   4:1 square/square
contraction, {\it International symposium on Applications of Laser
Techniques to Fluid Mechanics}, Lisbon, Portugal (2004).

[13]  Purmode, B. and Crochet, M. J., Flows of polymer solutions
through contractions Part I: Flows of polyacrylamide solutions
through planar contractions,  {\it J. Non-Newtonian Fluid Mech.},
{\bf 65} (1996) 269-289.

[14]  Genieser L. H., Armstrong, R.C., and Brown R. A.,Comparison of
measured centerline stress and velocity fields with predictions of
viscoelastic constitutive models, {\it J. Rheol.}, {\bf 47} (6)
(2003) 1331-1350.

[15]  Joseph, D.D., Fluid Dynamics of Viscoelastic Liquids,
 {\it Springer-Verlag},
  New York (1990).

[16]  Moffat, H.K., Viscous and Resistive Eddies Near Sharp Corners,
 {\it J.
 Fluid Mech.}, {\bf  18} (1964) 1-18.

[17]  Strogatz, S.H., Nonlinear Dynamics and Chaos,
 {\it Addison-Wesley},
 Reading, MA (1995).

[18]  Iooss, G. and Joseph, D. D., Elementary stability and
 Bifurcation theory,
 {\ it Springer Verlag}, New York (1980).

[19]  Guckenheimer, J. and Holmes, P., Non-Linear Oscillations,
 Dynamical
 Systems And Bifurcations of Vector Fields, {\it Springer Verlag},
 New York
 (1983).

[20]  Fietier, N. and Denville, M. O.,
 Time-Dependent algorithms for the simulation of viscoelastic flows
 with spectral element methods: Application and stability,
 {\it J. Com. Phys.},
 {\bf 186} (1)  (2003) 93-121.

[21]  Alves, M. A., Oliveira, P. J. and Pinho, F. T., Benchmark
solutions for the flow of Oldroyd-B and PTT fluids in planar
contractions,  {\it J. Non-Newtonian Fluid Mech.}, {\bf 110} (2003)
45-75.

[22]  Alves, M. A., Pinho, F. T. and Oliveira, P. J., Flow of
visco-elastic fluids past a cylinder: finite-volume high resolution
methods,  {\it J. Non-Newtonian Fluid Mech.}, {\bf  97} (2001)
207-232.

[23]  Phillips, T. N. and Williams, A. J., Viscoelastic flow through
a planar contraction using a semi-Lagrangian finite volume method,
{\it J. Non-Newtonian Fluid Mech.},  {\bf 87} (1999) 215-246.

[24]  Aboubacar, M. and Webster, M. F., A cell-vertex finite
volume/element method on triangles for abrupt contraction
viscoelastic flows, {\it J. Non-Newtonian Fluid Mech.}, {\bf 98}
(2001) 83-106.

[25]  Aboubacar, M., Matallah, H.  and Webster, M. F., Highly
elastic solutions for Oldroyd-B and Pan-Thien/Tanner fluids with a
finite volume/element method: Planar contraction flows, {\it J.
Non-Newtonian Fluid Mech.}, {\bf 103} (2002) 65-103.

\vfill\eject
\end{document}